\documentclass[pdflatex,sn-nature]{sn-jnl}

\newcommand{\Hila}[1]{\color{black}{#1}\color{black}}

\usepackage{graphicx}%
\usepackage{multirow}%
\usepackage{amsmath,amssymb,amsfonts}%
\usepackage{amsthm}%
\usepackage{mathrsfs}%
\usepackage{siunitx}

\usepackage{xcolor}%

\newcommand{\apj}{Astrophys. J.} 
\newcommand{\apjl}{Astrophys. J. Lett.} 
\newcommand{\apjs}{Astrophys. J. Suppl. Ser.} 
\newcommand{\aap}{Astron. Astrophys.} 
\newcommand{\mnras}{Mon. Not. R. Astron. Soc.} 
\newcommand{\nat}{Nature} 
\newcommand{\bain}{Bulletin of the Astronomical Institutes of the Netherlands} 

\newcommand{\sifig}[1]{Supplementary Figure \ref{#1} }
\newcommand{\sitab}[1]{Supplementary Table \ref{#1} }

\begin{document}

\title{The origin of hypervelocity white dwarfs in the merger-disruption of He-CO white dwarfs}







\author[1,*,+]{Hila Glanz}

\author[1,*,+]{Hagai B. Perets}

\author[2,3]{Aakash Bhat}

\author[4]{Ruediger Pakmor}

\affil[*]{glanz@campus.technion.ac.il; hperets@physics.technion.ac.il}
\affil[+]{these authors contributed equally to this work}


\affil[1]{Technion - Israel Institute of Technology \\
Haifa, 3200002, Israel}
\affil[2]{Institut für Physik und Astronomie, Universität Potsdam\\ Karl-Liebknecht-Str. 24/25, 14476 Potsdam-Golm, Germany}
\affil[3]{Dr. Karl-Remeis Observatory, Sternwartstr.7,
96049 Bamberg, Germany}
\affil[4]{Max-Planck-Institut für Astrophysik\\ 
Karl-Schwarzschild-Str. 1, D-85748 \\
Garching, Germany}

\abstract{
 Hypervelocity white dwarfs (HVWDs) are stellar remnants moving at speeds exceeding the Milky Way's escape velocity. The origins of the fastest HVWDs are enigmatic, with proposed formation scenarios facing challenges explaining both their extreme velocities and observed properties. Here we report a three-dimensional hydrodynamic simulation of a merger between two hybrid Helium-Carbon-Oxygen white dwarfs (HeCO WDs with masses of 0.69 and 0.62 M$_\odot$). We find that the merger leads to a partial disruption of the secondary WD, coupled with a double-detonation explosion of the primary WD. This launches the remnant core of the secondary WD at a speed of $~2000$ km s$^{-1}$, consistent with observed HVWDs. The low mass of the ejected remnant and its heating from the primary WD's ejecta explain the observed luminosities and temperatures of hot HVWDs, which are otherwise difficult to reconcile with previous models (such as the D6). This discovery establishes a new formation channel for HVWDs and points to a previously unrecognized pathway for producing peculiar Type Ia supernovae and faint explosive transients.
   
}

\maketitle




Hypervelocity white dwarfs (HVWDs \cite{Jor+12, Ven+17,Shen2018,Elbadry+23}) are enigmatic objects, hurtling through our galaxy at speeds exceeding the Milky Way's escape velocity\cite{Jor+12, Ven+17,Rad+18, Shen2018,Rad+19,Elbadry+23}. Their existence challenges our understanding of stellar evolution and the dynamics of close binary systems. While most white dwarfs exhibit velocities typical of the Galactic disk or halo, HVWDs possess extreme speeds, implying a violent ejection mechanism. The advent of high-precision astrometry from the Gaia mission \cite{Gaia2018, Gaia2021} has led to the identification of seven HVWD candidates with velocities exceeding 1000 km/s \cite{Shen2018, Elbadry+23}, further deepening the mystery surrounding their origin. Of these, two of the first identified HVWDs (D6-2 and D6-3) have very low measured radial velocities (RVs), possibly inconsistent with the far higher measured tangential velocities, casting doubt on their identification as HVWDs and/or their measured velocities \cite{Shen2018,2023IgoshevHypervelocity,Sch+24}, while the other five have large radial velocities consistent with typical $\sim2000$ km s$^{-1}$ (see Supplementary Table 1). As they had more time to be excited by galactic perturbations, old WDs might also have non-negligible peculiar velocities of up to a few hundred km s$^{-1}$, suggesting a larger uncertainty in ejection velocity. In particular, in-going (towards the direction of the Galactic disk) HVWDs could not have originated from the disk, suggesting large ages and peculiar velocities for their progenitors at the time of ejection. D6-2 and J0927 are in-going HVWDs, which might suggest that the inferred extreme ejection velocity of J0927 (the highest velocity inferred; $2519^{+271}_{-147}$ \cite{Elbadry+23}) could be lower, more in par with other HVWDs, and potentially closer to  
$\sim2000$ km s$^{-1}$, but conversely, could also be higher.   

 Lower velocity HVWDs with velocities of up to hundreds of km s$^{-1}$ were suggested to be produced from partial deflagration supernovae (SNe) of WDs and indeed, such objects were later identified; see \cite{Jor+12,Ven+17,Elbadry+23, Bha+24,Won+24} and references therein. \cite{2003HansenTypeIaWithHVWD} suggested the possibility of ejecting high-velocity evolved stars with velocities up to 100 km s$^{-1}$ in the context of the single degenerate scenario for type Ia SNe, i.e., a \Hila{Blaauw}-kick scenario for companions of exploding WDs\cite{1961BlaauwLick}).
A leading hypothesis for the formation of bona-fide HVWDs with $>1000$ km s$^{-1}$ follows a similar direction of a \Hila{Blaauw}-kick following a SN in a binary system, but for a different model for type Ia SNe, in which the ejected companion is a WD. It involves the "dynamically driven double-degenerate double-detonation" (D6) scenario \cite{Gui+10, Pakmor2013, Sato+15, Shen2018} (based on earlier double detonation scenarios not involving two WDs \cite{Ibe85, Iben+87}). In this model, a close binary system consisting of two white dwarfs (one typically massive $>0.85\, M_\odot$) undergoes a gravitational wave inspiral. As the primary white dwarf accretes a small amount of Helium from its companion, a Helium detonation is triggered in the accreted and any low-mass pre-existing Helium shell, leading to a second, carbon detonation in the core and a Type Ia SN explosion. The companion white dwarf, no longer gravitationally bound, is flung outward at high speed. As we discuss in detail below, while we also consider the case of a merger of two WDs, we focus on low-mass WDs, which show a different behavior, and our findings differ significantly from the D6 model in this context.

Nevertheless, the D6 scenario faces difficulties in explaining both the extreme HVWD velocities ($>$ 2000 km s$^{-1}$) and the observed properties of these objects\cite{Bau+21}. Recent studies \cite{Bha+24} have shown that detonation of a massive WD primary scenario as envisioned in the D6 scenario does not produce the high temperatures and inflated radii on observable timescales in some HVWDs, particularly for those with the highest velocities, for which the D6 scenario requires high mass HVWDs ($>1$ M$_\odot$\cite{bauer2019,Bau+21}). 
\Hila{Moreover, the predicted rate of HVWD formation from D6 events, if this were the primary channel for the origin of Type Ia supernovae, appears to be significantly higher than the observed rate\cite{Shen2018, 2023IgoshevHypervelocity}. Additionally, HVWDs have not been found in close proximity to \emph{secure} Type Ia supernova remnants\cite{2023IgoshevHypervelocity,Shi+23}, although a possible association has been noted for one case\cite{Shen2018}.}{}
More recent studies suggested that companions might also explode\cite{Tan+19,2021PakmorZenatiHybridIgnition,2022Pakmor,2024Boos}, which could potentially resolve the rates and supernova remnant issue, without significantly affecting the SN observations \cite{2022Pakmor, 2024Pollin,2024Boos}. However, in this case, no HVWD would form at all, requiring an alternative origin, like the one suggested here.

Here, we present a novel scenario for the origin of HVWDs based on a three-dimensional hydrodynamical simulation of a merger between two hybrid Helium-Carbon-Oxygen white dwarfs (HeCO WDs). These WDs are likely the remnants of previously interacting stars that underwent a common-envelope (CE) phase, resulting in a thick He shell surrounding a degenerate CO core (see \cite{Ibe+85,Zenati2019} and references therein). Our simulation reveals a previously unexplored pathway for HVWD formation, distinct from the standard D6 model. We find that the merger of two HeCO WDs (0.68 and 0.62 M$_\odot$) can lead to a partial tidal disruption of the lower-mass white dwarf (i.e. the secondary or companion), coupled with a double-detonation explosion of the primary. This process launches the remnant \emph{low-mass} hot core of the disrupted white dwarf at hypervelocity speeds, which can remain inflated and hot for long timescales, consistent with observations. Thereby, our findings provide a compelling explanation for the origin of the fastest HVWDs and offer new insights into the diversity of explosive stellar phenomena.

\begin{figure}
\includegraphics[width=\linewidth]{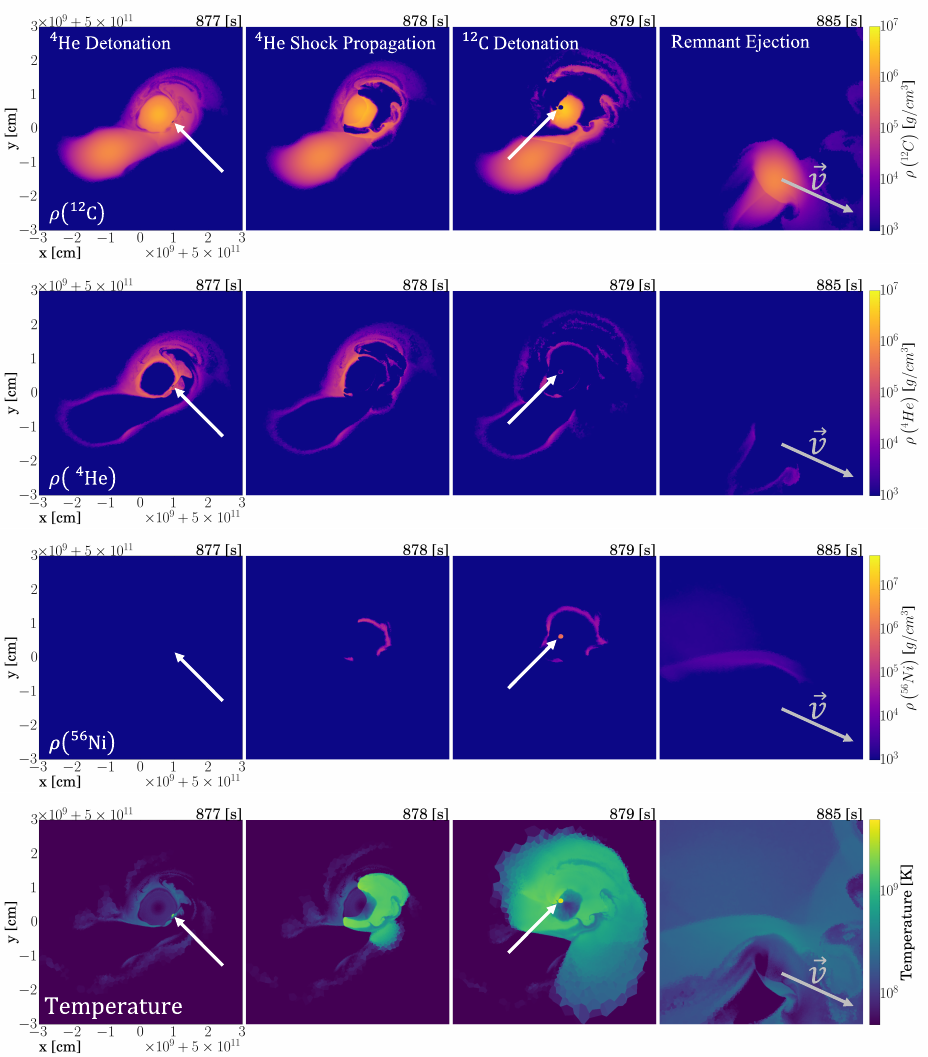}
\caption{\textbf{WD disruption and shock Propagation.} The panels show the time evolution from the time of the ignition of the Helium detonation (left panels) to the time when the shock converges in the CO core of the primary WD (third from the left), and finally the ejected remnant in the right panels. White arrows point to the detonation points, gray arrows indicate the remnant kick velocity.}
\label{fig:double_detonation}
\end{figure}

We model the merger using 3D hydrodynamical simulation using the \texttt{Arepo} code\cite{2010SpringelArepo} (see Methods). Figure \ref{fig:double_detonation} illustrates the key stages of the merger and explosion. A full movie showing the merger can be found in the SI (Supplementary Video 1), as well as a figure with more detailed snapshots (Sect. 2.2). As the secondary WD approaches the primary, it is tidally deformed and begins to transfer mass. This mass transfer eventually triggers a helium detonation on the surface of the primary WD. In the D6 scenario the high temperature and density conditions on the massive primary lead to early Helium detonation, occurring while the secondary is still intact, and after minimal accretion from the companion. However, in the low-mass HeCO cases shown here, a Helium detonati   on occurs only after the secondary has already begun to be significantly disrupted. In particular, a significant mass of Helium has already been transferred from the secondary to the primary, allowing for stronger, more energetic Helium detonation. The Helium detonation propagates around the primary, converging on the opposite side and driving a shock wave into the core. This shock wave initiates a secondary detonation in the Carbon-Oxygen core, leading to a supernova explosion, leaving the secondary unbound and ejected at a higher velocity. The very close approach and partial disruption allow for higher orbital (and eventually ejection) velocities and stronger heating of the companion. We note that since we followed the evolution for 200s, and the relatively low mass and density remnant has not been detonated, we don't expect it to ignite, as it further cools down in the long term, nor do we find evidence for this in our models of the long-term evolution discussed below.

Although the secondary WD is partially disrupted during the merger (stripped of $0.13$ M$_\odot$, and transferred $0.013$ M$_\odot$ to the primary), its core remains intact, when the Carbon-Oxygen detonation ensues. This core, tidally and debris-impact-heated and now unbound, is ejected following the supernova explosion at a velocity of 2061 km s$^{-1}$. The properties of the ejected remnant are summarized in Table \ref{tab:remnant_properties}.  The remnant has a mass of 0.492 M$_\odot$, including a small amount of Helium (0.003 M$_\odot$) and traces of heavier elements synthesized during the explosion. Due to the strong tidal interaction during the merger, the ejected HVWD is expected to be rapidly rotating. The initial rotation velocity should be comparable to the ejection velocity. However, as the remnant expands and cools, its rotation velocity decreases due to angular momentum conservation. At the final snapshot of our simulation, the HVWD has a total angular momentum of $2.48\times10^{50}$ g cm$^2$ s$^{-1}$; during its long-term evolution, the WD expands and its rotational velocity is expected to be significantly lower. For the currently inferred radius of $\sim0.05$ R$_\odot$ for some of the HVWDs, we would expect a rotation velocity of $\sim 770$ km s$^{-1}$.

\begin{table}
\centering
\begin{tabular}{c|c|c|c|c|c}
$M_{\text{b,Tot}}$ & $M_{\text{b,He}}$ & $M_{\text{b,IGE}}$ & $M_{\text{b,IME}}$ & Kick Velocity & Angular Momentum\\
$[\text{M}_\odot]$ & $[\text{M}_\odot]$ & $[\text{M}_\odot]$ & $[\text{M}_\odot]$ & [km s$^{-1}$] & [g cm$^2$ s$^{-1}$]\\
\hline
0.49 & $3.1\cdot{10}^{-3}$ & $3\cdot {10}^{-3}$ & $1\cdot {10}^{-2}$ & 2061 & $2.48\times10^{50}$
\end{tabular}
\caption{\textbf{Final properties of the remnant object (bound mass).} M$_b$ stands for bound mass, the second, column is the total mass of He in the remnant, the third column is the total mass of the elements we consider as an iron group (Scandium to Nickel), and the forth is the total bound mass of the intermediate elements (Fluorine to Calcium). Bound material is calculated by finding all mass with a positive energy, including the internal energy. The velocity refers to the velocity of the highest central density region of the remnant. The last column shows the total angular momentum in the remnant.}
\label{tab:remnant_properties}
\end{table}

The supernova explosion resulting from the double detonation releases $1.13 \times 10^{51}$ erg of nuclear energy (Table \ref{tab:SN_properties}). The explosion ejects a total of 0.82 M$_\odot$ of material, including 0.1  M$_\odot$ of iron-group elements, of which 0.072 M$_\odot$ is radioactive $^{56}$Ni, and 0.42 M$_\odot$ of intermediate-mass elements.
Interestingly, the Helium shell burning and detonation prior to the CO detonation already produces a few 0.01 M$_\odot$ of both radioactive $^{56}$Ni, as well as faster decaying radioactive elements $^{48}$Cr and $^{52}$Fe which could affect the early light-curve, and typically not expected in normal type Ia SNe (see the radioactive elements production evolution in time in Figure \ref{fig:radioactive} and the final composition in Supplementary Table 2). This finding agrees with previous He-detonation studies \cite{2011Waldman,2012MNRASSim,2023Zenati} with similar element production in their corresponding cases, even though our heavier 0.59 M$_\odot$ CO core has less He before the first detonation (about 0.13 M$_\odot$), (see Supplementary Table 3 for the element produced in the first (He) detonation,).  Comparing the total element production with previous works, we find that despite the differences in total mass and He shells, as well as differences in the numerical codes, our results also lie in the same regime as in the more similar double-detonation cases of Sim et al. 2010\cite{2012MNRASSim} (see Supplementary Table 2).  We note that it is still unclear what the conditions result in double detonation vs. a single He-shell detonation. Generally, double-detonation is expected to occur when both the He-shell and the CO core are denser, typically corresponding to higher WD mass, and when the He detonation is stronger, and produces a stronger shock into the CO core.

\begin{figure}
\includegraphics[width=\linewidth]{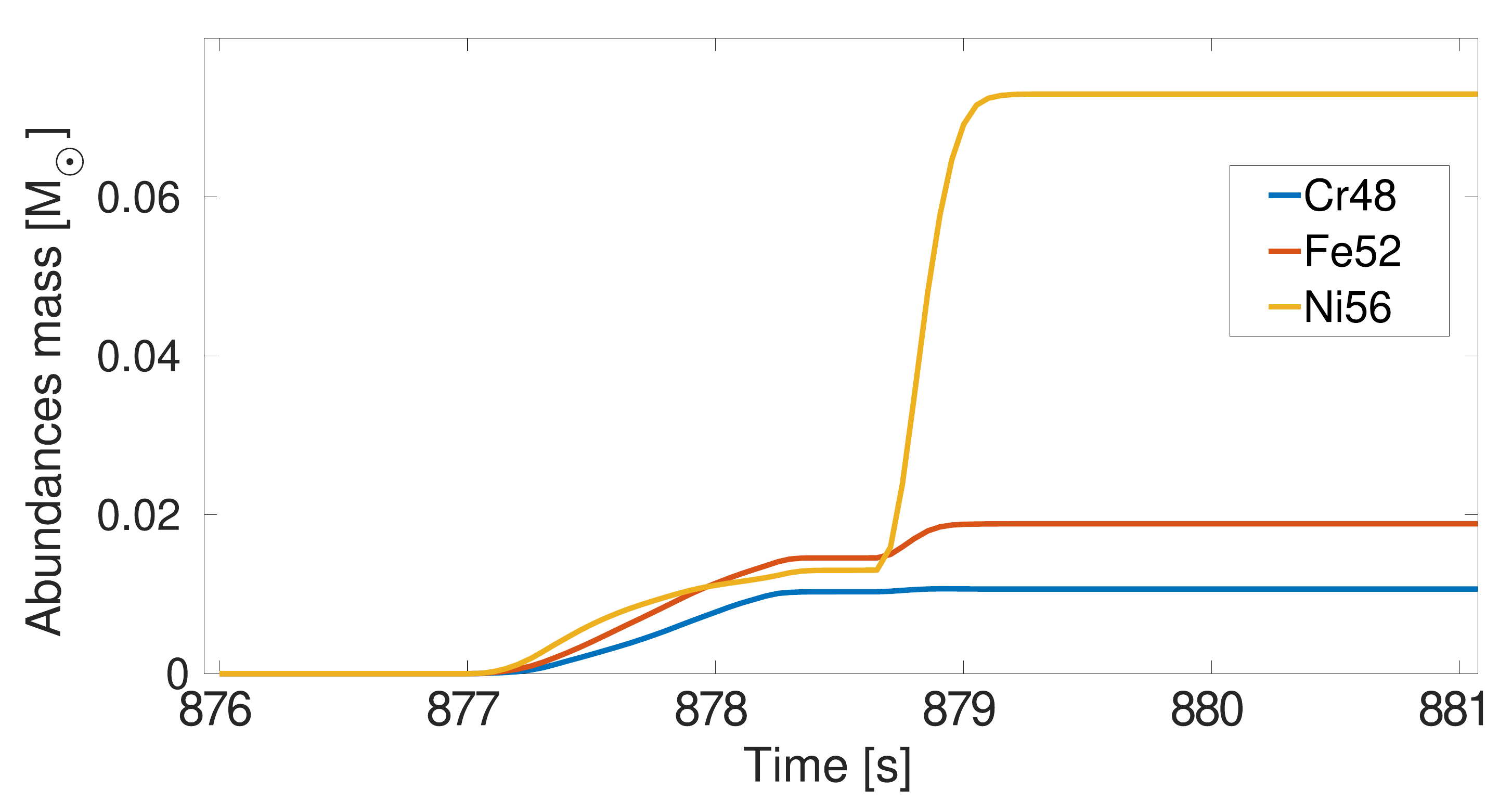}
 \caption{\textbf{Production of radioactive elements throughout the merger.} The early Helium burning and detonation already produces a significant amount of radioactive elements, with a few 0.01 M$_\odot$ of radioactive $^{56}$Ni, as well as faster decaying radioactive elements $^{48}$Cr and $^{52}$Fe produced at this stage. The bulk of the $^{56}$Ni is produced by the CO detonation beginning at t=878.8 s.}
 \label{fig:radioactive}
\end{figure}

\begin{table}
\centering
\begin{tabular}{c|c|c|c|c|c|c}
Total Nuclear Energy & $M_{\text{u,}1}$ & $M_{\text{u,}2}$ & $M_{\text{u,Tot}}$ & $M_{\text{u,IGE}}$ & $M_{\text{u,IME}}$ & SN CoM Velocity \\
  $[$erg] & $[\text{M}_\odot]$ & $[\text{M}_\odot]$ & $[\text{M}_\odot]$ & $[\text{M}_\odot]$ & [$\text{M}_\odot]$ & [km s$^{-1}$]\\
\hline
$1.126\cdot{10}^{51}$  & 0.68 & 0.14 & 0.82 & 0.1 & 0.42 & 1187
\end{tabular}
\caption{\textbf{Final properties of the supernova (the ejected, unbound material).} $M_u$ stands for the total unbound mass, 1 stands for the ejected mass from the primary, and 2 for the secondary. IGE are Iron group elements (Scandium to Nickel), and IME are Intermediate elements (Fluorine to Calcium). SN velocity is the unbound mass's center of mass velocity (CoM).}
\label{tab:SN_properties}
\end{table}

Table \ref{tab:SN_properties} summarizes the SN explosion properties. Given the low $^{56}$Ni produced, the low ejecta velocity (median 7000-8000 km s$^{-1}$),, and the relatively small amount of ejecta, the SN is likely to be faint and peculiar in respect to normal type Ia SNe. These properties (low mass, low $^{56}$Ni and low ejecta velocity)might be consistent with 91bg-like or 02es-like SNe \cite{2017bookTaubenberger}. The low abundances of radioactive elements may give rise to low photospheric temperatures, and together with the large  ${}^{44}$Ti abundance produced in the outer ejecta (see Supplementary Table 3 and Supplementary Table 2) could potentially induce early prominent Ti lines in the spectra, consistent with 91bg-like SNe. The properties of this SN might also resemble some properties of a Ca-rich SN, e.g. large abundances of intermediate elements such as ${}^{44}$Ti and ${}^{40}$Ca (see Supplementary Table 2). However, the ${}^{56}$Ni mass produced, and the total energetics are a few times higher than typical Ca-rich SNe. We do note that all of these sub-types of SNe are preferentially observed in old environments in early-type galaxies \cite{Per+10, 2017bookTaubenberger}. In addition, since the primary also had a large orbital velocity at the time of the explosion, the SN produced should have a large center-of-mass velocity. This could potentially be identified as a shift in the velocity measurement of nebular spectra taken long after the explosion of such a peculiar, faint SN. The supernova remnants (SNR) of such SNe, if identified, should also have such a shift. In particular, if paired with an identified HVWD, it should have a velocity in a counter direction to that of the HVWD. We find the SNR center of mass velocity in our simulation to be  1187 km s$^{-1}$.

\begin{figure}
\includegraphics[width=\linewidth]{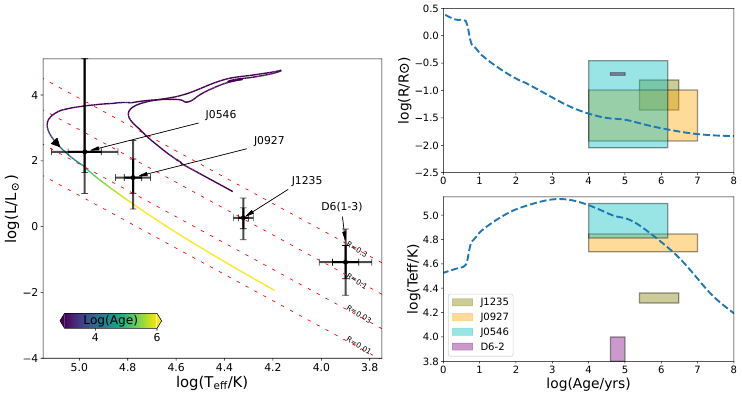}
 \caption{\textbf{Evolution of the MESA model and comparison with HVWD observations. } Left: The HR diagram for our model compared with the observed values. The error bars depict the 1 and 2 $\sigma$ uncertainties. The black arrow marks the point when the star is $10^4$ years old. Right: The radius and temperature evolution of the model as a function of its age. In particular, the HVWDs J0546, and J0927 may be explained by our model, within the 1-sigma (J0546) and $2\sigma$ (J0927; or 1$\sigma$ too, if ref. \cite{Elbadry+23} inferred temperature is adapted) uncertainties. See further discussion in the SI regarding J1332 and its omission here 2.4}
 \label{fig:HR}
\end{figure}

Following the simulation, we record the remnant properties (densities, compositions, temperature) found in the simulation (see Table \ref{tab:remnant_properties}),  and then follow its long-term evolution, following the same approach as ref. \cite{Bha+24}, by mapping the properties into 1D and then using the \texttt{MESA} stellar evolution code\cite{2011MESA} (see Methods). 
Figure \ref{fig:HR} shows the long-term evolution of the ejected HVWD in the Hertzsprung-Russell diagram (HR diagram). The remnant is initially hot and luminous, and experiences some super-Eddington mass-loss, but then cools and fades over time as it expands and releases its thermal energy. The low mass of the remnant contributes to its inflated state and prolonged cooling time.

The observed physical properties of several HVWD candidates are also plotted in Figure \ref{fig:HR}, taken from \cite{Elbadry+23}, besides for J0927\cite{Wer+24}. The radius and temperature of our simulated remnant align well with the observed values, particularly for the HVWD J0546 and J1332, and to a lesser extent with J0927 (consistent within 2-sigma uncertainty, or 1$\sigma$, if ref. \cite{Elbadry+23} inferred temperature is adapted), but see further discussion on the properties J1332 in the SI, section 2.4. This suggests that our proposed formation scenario can explain the observed properties of the more reliable and highest-velocity HVWDs. That being said, D6-1 and J1235 do show consistent reliable high radial velocities but are not directly explained by our specific simulation. Nevertheless, given the large phase space of HeCO WD mergers, it is possible that more consistent models can be found with more/less massive HeCOs configurations (see also discussion on velocities in the SI, section 2.3).

Our 3D hydrodynamical simulation has revealed a new channel for the formation of HVWDs through the merger of two hybrid HeCO white dwarfs. This scenario differs significantly from the previously proposed D6 model, which involves the detonation of a \emph{massive} CO white dwarf, leading to a normal type Ia SN, following very little mass transfer, leaving the companion intact. In our simulation, the merger of two lower-mass HeCO WDs leads to a partial tidal disruption of the secondary WD and a double-detonation explosion of the low-mass primary. The close approach of the binary down to the point of partial tidal disruption allows for the larger velocity of the companion, even though the primary is relatively low-mass. Hence, this leads to the ejection of the remnant core of the secondary as an HVWD with a velocity exceeding 2000 km s$^{-1}$. This leads to a fainter SN than a normal type Ia.

This novel channel offers several advantages in explaining the observed properties of HVWDs. First, it naturally produces the high ejection velocities observed for some HVWDs, which are difficult to achieve in the D6 scenario without invoking massive progenitor WDs, which are rare\cite{bauer2019,Bau+21}. Second, the low mass of the ejected remnant and its heating from the primary WD's ejecta explain the observed luminosities and inflated radii of HVWDs, which can not be reconciled with the D6 model for HVWDs with such velocities.

Our findings have broader implications for our understanding of stellar evolution and explosive transients. They highlight the potential importance of HeCO WDs in producing a variety of thermonuclear events. The merger of HeCO WDs may lead to peculiar Type Ia supernovae with fainter luminosities and potentially distinct observational signatures compared to standard SNe Ia. Further studies are needed to characterize the detailed properties of these explosions and their potential contribution to the observed diversity of SNe Ia.

The discovery of HVWDs was initially hailed as a smoking-gun signature of the D6 scenario and its potential role in producing the majority of SNe Ia. However, our results suggest that HVWDs may originate from a wider range of progenitor systems and explosion mechanisms. Together with the various other challenges to the D6 scenario discussed above (e.g., rates and velocities, no evidence for HVWD inside type Ia SNRs), our results cast further doubt on the D6 scenario as the \emph{dominant} channel for SNe Ia \Hila{with ejected HVWDs,}{} and emphasize the need to explore alternative pathways.

The current study is limited by the computational cost of 3D hydrodynamical simulations, which restricts the exploration of a wider parameter space of HeCO WD mergers. Future studies with larger parameter surveys will help to constrain the range of HVWD properties and the rates of these events. Nevertheless, our findings demonstrate the viability of HeCO WD mergers as a source of HVWDs and open up new avenues for research into the explosive fates of binary white dwarf systems (see also \cite{Per+19,2021PakmorZenatiHybridIgnition,2023Zenati} for other explosive transients involving HeCO WDs). Moreover, analytic estimates (see SI; section 2.3) suggest that it is likely that the phase space of merging double HeCO WDs could produce HVWDs of a wider range of velocities and masses beyond the specific point derived from the fully modeled case shown here. It could naturally explain (and predict) a wider range of velocities in the range $2000\pm300$ km s$^{-1}$ expected for HVWDs from the partial HeCO tidal-disruption channel. Indeed, the inferred velocities (within the $1-\sigma$ uncertainties) of all besides D6-2 (which is the least reliable given its very low radial velocity) of the HVWDs reside in this range. In fact, these might explain the preference for this velocity range. 

In summary, we have demonstrated that the merger of two HeCO WDs can produce HVWDs with properties consistent with observations, both in terms of ejection velocities, ages, and physical properties. This novel channel provides a compelling explanation for the origin of the fastest HVWDs and sheds new light on the diversity of explosive transients in the Universe.

\section{Methods}
\label{sec:Method}
We investigate the formation of HVWDs through the merger of two hybrid Helium-Carbon-Oxygen (HeCO) white dwarfs using a three-dimensional magnetohydrodynamical simulation. This is the first study to explore this scenario, in particular, in 3D, allowing us to capture the complex dynamics of the merger process and the subsequent ejection of the HVWD.

\subsection{HeCO White Dwarfs}
\label{Methods-HeCO-WDs}
HeCO white dwarfs are a class of stellar remnants characterized by a Helium-rich outer layer surrounding a Carbon-Oxygen core. Their formation is linked to binary star interactions, where processes like Roche lobe overflow and common envelope evolution lead to the accumulation of Helium on the white dwarf \cite{Ibe+85, Zen+18,2021PakmorZenatiHybridIgnition}. These hybrid WDs can have masses and compositions that fall between those of typical Helium WDs and Carbon-Oxygen WDs, making their merger outcomes particularly interesting.

More specifically, we investigate the fate of a double degenerate binary system consisting of a primary Helium-rich HeCO WD, with the total mass of $0.69\,\mathrm{M_\odot}$ made of a CO core of $0.59\,\mathrm{M_\odot}$ and a massive Helium shell of $0.1\,\mathrm{M_\odot}$ (see details in \cite{2021PakmorZenatiHybridIgnition}), and a secondary hybrid HeCO WD with total mass of $0.62\,\mathrm{M_\odot}$ made of a CO core of $0.58\,\mathrm{M_\odot}$ and a massive Helium shell of $0.038\,\mathrm{M_\odot}$ (following \cite{Zen+19a}). See Sec. 2.5 for a discussion on the rate of such interactions. Note that the much higher He shell for this hybrid is attained through a somewhat different evolution than that described in \cite{Zenati2019}, as explored in \cite{2021PakmorZenatiHybridIgnition}.

For the density profile and composition of the $0.62\,\mathrm{M_\odot}$ WD, we use a similar method as in \cite{2022Pakmor}, where we generate the 1D profile in hydrostatic equilibrium, with 0.0365M$_\odot$ He shell, following the results of \cite{Zen+18}.

\subsection{Hydrodynamical Simulation}

We model the merger of two HeCO white dwarfs using the moving-mesh hydrodynamics code \texttt{Arepo} \cite{2010SpringelArepo,2011PakmorArepoMHD,2015ZhuMHDArepoWD,2020WeinbergArepoPublic}. This code provides a self-gravitating, adaptive mesh framework for simulating astrophysical phenomena, incorporating a Helmholtz equation of state for degenerate matter and a 55-isotopes nuclear reaction network to capture the thermonuclear processes involved in the merger. A detailed description of WD merger simulations in \texttt{Arepo} can be found in \cite{2021PakmorZenatiHybridIgnition}, and our models closely follow similar approaches and modeling. We run our simulation with a cell mass resolution of $9\times {10}^{26}$g, a minimum cell volume of $\sim {10}^{19}$cm$^3$,  $\sim 3.5 \times {10}^6$ cells, in a total box size of ${10}^{12}$cm to incorporate the explosion and the expansion of the remnant.

\subsubsection{Initial Conditions}

The initial conditions for our simulation are derived from 1D stellar evolution models. These models provide the density, temperature, and composition profiles for both the primary and secondary HeCO white dwarfs.  The specific parameters of our simulated binary system are:

{\bf Primary WD:} Mass = 0.69 M$_\odot$, CO core = 0.59 M$_\odot$ (0.26 M$_\odot$ ${}^{12}$C and 0.33 ${}^{16}$O), He shell = 0.1 M$_\odot$ as modeled in \cite{2021PakmorZenatiHybridIgnition} using \texttt{MESA} \cite{2011MESA, Paxton2013, Paxton2019}, and mapped to 3D.
{\bf Secondary WD:} Mass = 0.62 M$_\odot$, CO core = 0.58 M$_\odot$ (equal amount of ${}^{12}$C and ${}^{16}$O), He shell = 0.038,  M$_\odot$, where the composition is based on \texttt{MESA} models in \cite{Zen+18}. This model was generated synthetically in the same method as in \cite{2022Pakmor}.

The 3D white dwarf models are initialized in \texttt{Arepo}, relaxed to ensure stability, and then placed in a binary configuration within a large simulation domain. The initial separation of the binary is chosen to be wide enough to minimize initial tidal forces, $5.1 \times 10^4$ km, and the corresponding orbital period was $177.5$s.

\subsubsection{Inspiral and Merger}
We simulate the inspiral of the binary system, initially accelerating it by artificially removing angular momentum to mimic the effect of gravitational wave emission. This artificial inspiral is halted just before the secondary WD enters the Roche radius of the primary and mass transfer becomes dynamically significant (see \cite{2021PakmorZenatiHybridIgnition} for a similar approach and further details). The binary separation at this point was $2.37\times{10}^4$km, with an orbital period of $8.7$s.

We then follow the fully realistic evolution of the merger. As the secondary WD approaches the primary, it is tidally deformed and disrupted ($\sim 100$s after the end of the accelerated inspiral stage). The dynamical mass transfer of Helium from the secondary triggers a Helium detonation on the surface of the primary. This detonation propagates around the primary, leading to a secondary detonation in the Carbon-Oxygen core and a supernova explosion. The partially disrupted secondary WD, still containing its core, is ejected at high velocity by the force of the explosion. These stages are presented in Supplementary Figure 1, while some are also in Figure \ref{fig:double_detonation} of the main part. A full movie of the simulation can be found in the SI (Supplementary Video 1).

\subsection{Evolution of the Bound Remnant HVWD}
To model the long-term evolution of the runaway WD we follow a similar approach as in \cite{Bha+24}. We map the output of the simulation ($\sim 300$ s after the explosion) to the 1D stellar evolution software \texttt{MESA}. This procedure has been applied before for \texttt{Athena++} simulations of Helium subdwarfs\cite{bauer2019} and more recently of Helium white dwarfs\cite{Won+24}, as well as \texttt{Arepo} simulations of massive white dwarfs \cite{Bha+24}. Here, we follow the procedure outlined in reference \cite{Bha+24}. The 3D profiles of the runaway WD (taken at the last snapshot, which is 10-20 times dynamical timescales after the explosion, allowing sufficient time for the WD to adjust) are averaged to give a 1D profile with temperature, density, and mass fraction information. A 22 isotope network approx\_21\_plusco56 is used to model the nuclear reactions during this evolution and includes the isotopes $^{1}$H, $^{3,4}$He, $^{12}$C, $^{14}$N, $^{16}$O, $^{20}$Ne, $^{24}$Mg, $^{28}$Si, $^{32}$S, $^{36}$Ar, $^{40}$Ca, $^{44}$Ti, $^{48,56}$Cr, $^{52,54,56}$Fe, $^{56}$Co, and $^{56}$Ni. It therefore includes the most important reactions of nickel decay and $\alpha$ capture on Carbon. We relax a white dwarf to our output profiles of composition and entropy, as outlined in \cite{Bha+24}, and we get rid of almost all of the outer layers where $^{56}$Ni dominates the atmosphere to such an extent that the material will become unbound on short timescales (but longer than the hydrodynamical simulation). This cut is chosen again to be the mass coordinate where the energy due to $^{56}$Ni decay is twice the gravitational binding energy of the star. This results in a final mass of 0.482 M$_\odot$, with only 0.005 M$_\odot$ cut-off. Once relaxed the white dwarf is allowed to evolve for a maximum time of 100 Myr. We present the results in Figure \ref{fig:HR} and Supplementary Figure 2. In addition, as we show in Supplementary Table 2, there is still bound He at the end of our simulation, which is kept there also when following its longer evolution; it might however, keep changing on an even longer timescale. Nevertheless, the existence of a bound He in the remnant will likely lead to He detection in the remnant's atmosphere.

\section*{Data Availability}
\Hila{The \texttt{MESA} input files and inlists for relaxation, heating, and evolution can be accessed at Zenodo \cite{Glanz2025Zenodo}.\\
All data underlying this research, which is not in Zenodo (in particular, the many too large, hydrodynamical simulation snapshots), will be available upon reasonable request to the corresponding authors.}{}

\section*{Code Availability}
We make use of the following codes: \texttt{Arepo}\cite{2010SpringelArepo,2020WeinbergArepoPublic,2011PakmorArepoMHD,pak+16}, \texttt{MESA}\cite{2011MESA,Paxton2013,Paxton2019,2Jer+23}, \texttt{Matplotlib} \cite{Hunter:2007}, \texttt{NumPy} \cite{harris2020arrayNUMPY}, and
\texttt{SciPy}\cite{2020SciPy-NMeth}. \texttt{MESA} inlists are available through the \texttt{MESA} community on Zenodo.

\section*{Acknowledgments}
HBP acknowledges support for this project from the European Union's Horizon 2020 research and innovation program under grant agreement No 865932-ERC-SNeX. HG acknowledges support for the project from the Council for Higher Education of Israel.  A.B.
was supported by the Deutsche Forschungsgemeinschaft (DFG)
through grant GE2506/18-1 and by the Kavli Summer Program which took place at MPA in Garching in July 2023, and was supported by the
Kavli Foundation. We would like to thank Evan Bauer and Robert Fisher for their valuable comments and discussions.

\section*{Author contributions statement}
HG led the key research, ran the hydrodynamical simulations, and analyzed their results. HBP initiated and supervised the project, suggested the main ideas, analyzed the results, and wrote the main parts of the paper. AB made the long-term WD evolution modeling and analyzed them. RP assisted with running the hydrodynamical simulations. All authors contributed to writing the paper, making the figures, and reviewing the manuscript. 

\textbf{Corresponding authors: } Correspondence and requests for materials should be addressed to Hila Glanz or Hagai Perets.

\section*{Competing interests statement} 
The authors declare no competing interests.

\newpage
\renewcommand\thefigure{\arabic{figure}}    
\setcounter{figure}{0}    

\renewcommand\thetable{\arabic{table}}    
\setcounter{table}{0}



%


\section{Supplementary Information}

\renewcommand{\figurename}{Supplementary Figure}
\renewcommand\thefigure{\arabic{figure}}    
\setcounter{figure}{0}

\renewcommand{\tablename}{Supplementary Table}
\renewcommand\thetable{\arabic{table}}    
\setcounter{table}{0}   

This supplementary information provides additional details and supporting material for the main text.

\subsection{The properties of the hypervelocity WDs}
\label{sec:SI-HVWDs}
\sitab{tab:hvwd-props} summarizes the physical and kinematic properties of the seven fastest hypervelocity WDs, from tables 2 and 3 in ref. \cite{Elbadry+23} (kinematics and radii) and from Refs. \cite{Shen2018,Cha+22,Wer+24,Wer+24b} for the temperatures and radii. We note here that for the cases of J0927 and J0546 the radius and distance values of \cite{Wer+24b} do not match those of \cite{Elbadry+23}. This is because the radii of the former are based on evolutionary tracks of white dwarfs, which might not hold for hypervelocity white dwarfs. \cite{Wer+24b} find a distance of $15^{+4}_{-8}$ kpc for J0546, which could make this star almost three times faster. Therefore, for these stars we use the spectroscopic temperature and $\log g$ from \cite{Wer+24b} but the radius from \cite{Elbadry+23}. 

\begin{table}
\centering
\begin{tabular}{l|c|c|c|c|c|c|c|c}
Name & $v_r$ & $v_\text{tot}$ & $v_\text{ejection}$ & $T_\text{eff}$ & $R$ & $d$ & $z$ & $t_\text{disk}$ \\
     & [km s$^{-1}$] & [km s$^{-1}$] & [km s$^{-1}$] & [K] & [R$_\odot$] & [kpc] & [kpc] & [Myr] \\
\hline
D6-1 & 1200$\pm$40 & 2045$^{+251}_{-190}$ & 2254$^{+248}_{-199}$ & 7000 & 0.23$^{+0.04}_{-0.03}$ & 1.91$^{+0.2}_{-0.2}$ & -0.57$^{+0.07}_{-0.06}$ & 0.63$^{+0.04}_{-0.04}$ \\
D6-2 & 80$\pm$10 & 1151$^{+59}_{-55}$ & 1051$^{+62}_{-52}$ & 7000 & 0.15$^{+0.01}_{-0.01}$ & 0.84$^{+0.05}_{-0.04}$ & -0.27$^{+0.01}_{-0.02}$ & -0.72$^{+0.01}_{-0.01}$ \\
D6-3 & -20$\pm$80 & 2248$^{+340}_{-324}$ & 2393$^{+377}_{-396}$ & 7000 & 0.17$^{+0.02}_{-0.02}$ & 2.26$^{+0.34}_{-0.35}$ & 0.93$^{+0.14}_{-0.14}$ & 2.24$^{+0.21}_{-0.17}$ \\
J1235-3752 & -1694$\pm$10 & 2670$^{+339}_{-250}$ & 2471$^{+351}_{-256}$ & 21000 & 0.104$^{+0.026}_{-0.030}$ & 4.53$^{+2.17}_{-1.82}$ & 1.73$^{+0.42}_{-0.50}$ & 1.87$^{+0.86}_{-0.25}$ \\
J0927-6335 & -2285$\pm$20 & 2753$^{+271}_{-148}$ & 2519$^{+271}_{-147}$ & 60000 & 0.052$^{+0.025}_{-0.020}$ & 4.07$^{+1.01}_{-0.70}$ & -0.70$^{+0.27}_{-0.35}$ & -1.32$^{+0.42}_{-0.42}$ \\
J0546+0836 & 1200$\pm$20 & 1699$^{+660}_{-390}$ & 1864$^{+682}_{-416}$ & 95000 & 0.051$^{+0.029}_{-0.021}$ & 4.02$^{+2.29}_{-1.67}$ & -0.69$^{+0.29}_{-0.40}$ & 0.61$^{+0.08}_{-0.08}$ \\
J1332-3541 & 1090$\pm$50 & 1464$^{+707}_{-331}$ & 1619$^{+707}_{-320}$ & 70000 & 0.017$^{+0.013}_{-0.007}$ & 1.63$^{+1.19}_{-0.68}$ & 0.74$^{+0.53}_{-0.30}$ & 0.64$^{+0.13}_{-0.14}$ \\
\hline
\end{tabular}
\caption{\textbf{Kinematic and physical properties of suspected D6 hypervelocity white dwarfs.} Columns show radial velocity, total velocity, ejection velocity, effective temperature, radius, distance, height above the Galactic plane, and flight time back to the Galactic disk.}
\label{tab:hvwd-props}
\end{table}

\begin{table}[h]
\centering
\begin{tabular}{c|c|c}
Element & Remnant Mass $[\text{M}_\odot]$ & Ejecta Mass $[\text{M}_\odot]$\\
\hline
${}^{}$n & \num{5.81e-23} & \num{2.67e-18}\\
${}^{}$H & \num{1.35e-15} & \num{3.69e-06}\\
${}^{4}$He & \num{3.15e-03} & \num{4.41e-02}\\
${}^{11}$B & \num{1.07e-19} & \num{1.01e-18}\\
${}^{12}$C & \num{2.37e-01} & \num{4.87e-02}\\
${}^{13}$C & \num{4.71e-10} & \num{4.26e-10}\\
${}^{13}$N & \num{1.11e-08} & \num{2.03e-07}\\
${}^{14}$N & \num{3.40e-08} & \num{1.92e-07}\\
${}^{15}$N & \num{1.63e-09} & \num{4.88e-08}\\
${}^{15}$O & \num{4.13e-08} & \num{3.55e-06}\\
${}^{16}$O & \num{2.38e-01} & \num{2.07e-01}\\
${}^{17}$O & \num{4.24e-10} & \num{3.36e-09}\\
${}^{18}$F & \num{6.58e-11} & \num{1.53e-09}\\
${}^{19}$Ne & \num{5.16e-10} & \num{1.38e-08}\\
${}^{20}$Ne & \num{3.81e-04} & \num{1.72e-02}\\
${}^{21}$Ne & \num{1.27e-09} & \num{3.72e-08}\\
${}^{22}$Ne & \num{1.09e-10} & \num{1.71e-09}\\
${}^{22}$Na & \num{2.60e-09} & \num{7.34e-08}\\
${}^{23}$Na & \num{5.06e-07} & \num{1.31e-05}\\
${}^{23}$Mg & \num{7.29e-07} & \num{3.64e-05}\\
${}^{24}$Mg & \num{3.19e-04} & \num{3.45e-02}\\
${}^{25}$Mg & \num{6.23e-08} & \num{2.84e-06}\\
${}^{26}$Mg & \num{5.04e-08} & \num{1.93e-06}\\
${}^{25}$Al & \num{1.87e-08} & \num{1.26e-05}\\
${}^{26}$Al & \num{1.75e-06} & \num{9.89e-05}\\
${}^{27}$Al & \num{6.45e-07} & \num{6.07e-05}\\
${}^{28}$Si & \num{4.57e-03} & \num{1.71e-01}\\
${}^{29}$Si & \num{4.03e-06} & \num{1.18e-04}\\
${}^{30}$Si & \num{8.33e-08} & \num{1.27e-05}\\
${}^{29}$P & \num{4.96e-06} & \num{2.83e-04}\\
${}^{30}$P & \num{5.70e-07} & \num{4.76e-05}\\
${}^{31}$P & \num{1.73e-06} & \num{1.77e-04}\\
${}^{31}$S & \num{9.54e-07} & \num{1.90e-04}\\
${}^{32}$S & \num{3.79e-03} & \num{1.19e-01}\\
${}^{33}$S & \num{1.15e-06} & \num{1.70e-04}\\
${}^{33}$Cl & \num{2.80e-08} & \num{7.03e-05}\\
${}^{34}$Cl & \num{2.12e-07} & \num{9.14e-05}\\
${}^{35}$Cl & \num{1.19e-08} & \num{1.50e-05}\\
${}^{36}$Ar & \num{8.96e-04} & \num{3.13e-02}\\
${}^{37}$Ar & \num{4.78e-07} & \num{3.09e-04}\\
${}^{38}$Ar & \num{2.20e-10} & \num{1.17e-07}\\
${}^{39}$Ar & \num{9.56e-14} & \num{3.32e-12}\\
${}^{39}$K & \num{8.66e-09} & \num{1.84e-05}\\
${}^{40}$Ca & \num{9.27e-04} & \num{4.40e-02}\\
${}^{43}$Sc & \num{3.23e-13} & \num{4.58e-09}\\
${}^{44}$Ti & \num{9.64e-07} & \num{5.07e-03}\\
${}^{47}$V & \num{1.02e-11} & \num{4.41e-07}\\
${}^{48}$Cr & \num{1.58e-05} & \num{1.06e-02}\\
${}^{51}$Mn & \num{3.07e-10} & \num{1.19e-06}\\
${}^{52}$Fe & \num{2.08e-04} & \num{1.94e-02}\\
${}^{56}$Fe & \num{7.03e-08} & \num{6.63e-05}\\
${}^{55}$Co & \num{1.54e-09} & \num{1.40e-06}\\
${}^{56}$Ni & \num{2.76e-03} & \num{7.02e-02}\\
${}^{58}$Ni & \num{1.09e-05} & \num{2.94e-04}\\
${}^{59}$Ni & \num{8.39e-09} & \num{3.39e-05}
\end{tabular}
\caption{\textbf{Final composition in the remnant (bound material) and the supernova (ejected, unbound material).}}
\label{tab:SN_elements}
\end{table}

\begin{table}[h]
\centering
\begin{tabular}{c|c|c}
Element & Remnant Mass $[\text{M}_\odot]$ & Ejecta Mass $[\text{M}_\odot]$\\
\hline
${}^{}$n & \num{1.03e-16} & \num{2.76e-14}\\
${}^{}$H & \num{1.47e-10} & \num{4.06e-06}\\
${}^{4}$He & \num{6.85e-03} & \num{4.11e-02}\\
${}^{11}$B & \num{4.10e-18} & \num{1.64e-18}\\
${}^{12}$C & \num{4.77e-01} & \num{1.27e-02}\\
${}^{13}$C & \num{1.78e-10} & \num{8.70e-11}\\
${}^{13}$N & \num{3.18e-08} & \num{2.03e-08}\\
${}^{14}$N & \num{7.03e-08} & \num{4.05e-08}\\
${}^{15}$N & \num{2.90e-10} & \num{6.07e-10}\\
${}^{15}$O & \num{1.61e-07} & \num{7.85e-07}\\
${}^{16}$O & \num{5.79e-01} & \num{4.70e-02}\\
${}^{17}$O & \num{1.47e-09} & \num{5.87e-10}\\
${}^{18}$F & \num{8.05e-10} & \num{2.99e-10}\\
${}^{19}$Ne & \num{3.54e-09} & \num{3.34e-09}\\
${}^{20}$Ne & \num{4.08e-03} & \num{7.06e-03}\\
${}^{21}$Ne & \num{1.37e-08} & \num{9.33e-09}\\
${}^{22}$Ne & \num{6.46e-10} & \num{3.61e-10}\\
${}^{22}$Na & \num{2.55e-08} & \num{1.84e-08}\\
${}^{23}$Na & \num{3.21e-06} & \num{1.72e-06}\\
${}^{23}$Mg & \num{1.01e-05} & \num{9.65e-06}\\
${}^{24}$Mg & \num{3.19e-03} & \num{1.11e-02}\\
${}^{25}$Mg & \num{6.31e-07} & \num{6.25e-07}\\
${}^{26}$Mg & \num{5.35e-07} & \num{2.84e-07}\\
${}^{25}$Al & \num{8.94e-08} & \num{7.60e-06}\\
${}^{26}$Al & \num{2.00e-05} & \num{2.13e-05}\\
${}^{27}$Al & \num{8.04e-06} & \num{2.63e-05}\\
${}^{28}$Si & \num{3.04e-03} & \num{2.62e-02}\\
${}^{29}$Si & \num{4.17e-06} & \num{9.88e-06}\\
${}^{30}$Si & \num{9.10e-07} & \num{8.48e-06}\\
${}^{29}$P & \num{2.61e-07} & \num{1.51e-04}\\
${}^{30}$P & \num{5.24e-06} & \num{9.23e-06}\\
${}^{31}$P & \num{5.02e-06} & \num{3.58e-05}\\
${}^{31}$S & \num{1.41e-05} & \num{1.35e-04}\\
${}^{32}$S & \num{5.56e-04} & \num{2.06e-02}\\
${}^{33}$S & \num{9.60e-06} & \num{3.96e-05}\\
${}^{33}$Cl & \num{2.59e-09} & \num{5.71e-05}\\
${}^{34}$Cl & \num{2.20e-06} & \num{7.13e-05}\\
${}^{35}$Cl & \num{1.79e-07} & \num{1.41e-05}\\
${}^{36}$Ar & \num{4.40e-05} & \num{9.40e-03}\\
${}^{37}$Ar & \num{2.22e-06} & \num{2.63e-04}\\
${}^{38}$Ar & \num{7.81e-10} & \num{1.22e-07}\\
${}^{39}$Ar & \num{4.12e-13} & \num{2.57e-12}\\
${}^{39}$K & \num{6.48e-08} & \num{1.86e-05}\\
${}^{40}$Ca & \num{1.01e-05} & \num{2.26e-02}\\
${}^{43}$Sc & \num{3.27e-12} & \num{5.12e-09}\\
${}^{44}$Ti & \num{5.31e-07} & \num{5.21e-03}\\
${}^{47}$V & \num{7.19e-11} & \num{5.69e-07}\\
${}^{48}$Cr & \num{1.92e-06} & \num{1.04e-02}\\
${}^{51}$Mn & \num{6.21e-10} & \num{1.40e-06}\\
${}^{52}$Fe & \num{2.87e-06} & \num{1.46e-02}\\
${}^{56}$Fe & \num{6.46e-09} & \num{6.69e-05}\\
${}^{55}$Co & \num{1.07e-09} & \num{1.56e-06}\\
${}^{56}$Ni & \num{2.38e-06} & \num{1.31e-02}\\
${}^{58}$Ni & \num{1.76e-08} & \num{2.29e-05}\\
${}^{59}$Ni & \num{6.15e-09} & \num{1.84e-05}
\end{tabular}
\caption{\textbf{Full composition after the first detonation ( at 878.5 s).} The second row is the element mass bound to the high density region, third row is the unbound mass.}
\label{tab:1st_detonation_elements}
\end{table}

\subsection{Detailed evolution of the merger: Movie and snapshots}
\label{sec:SI-movie}
A movie visualizing the merger of the two HeCO white dwarfs and the subsequent ejection of the HVWD is available online at \href{https://tinyurl.com/dd2hyperv}{https://tinyurl.com/dd2hyperv}. The movie shows the tidal disruption of the secondary WD, the double-detonation explosion of the primary, and the high-velocity ejection of the remnant core.

\begin{figure}
\includegraphics[width=\textwidth]{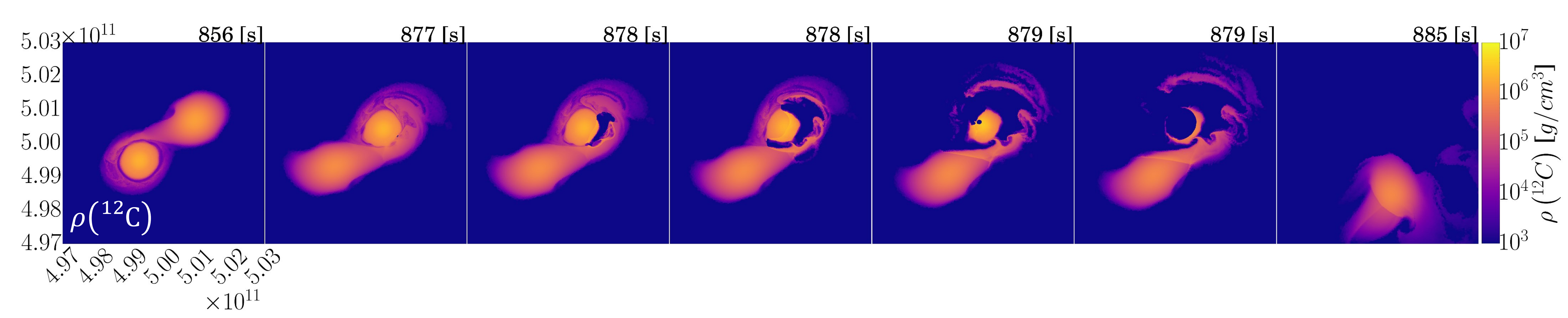}\\
\includegraphics[width=\textwidth]{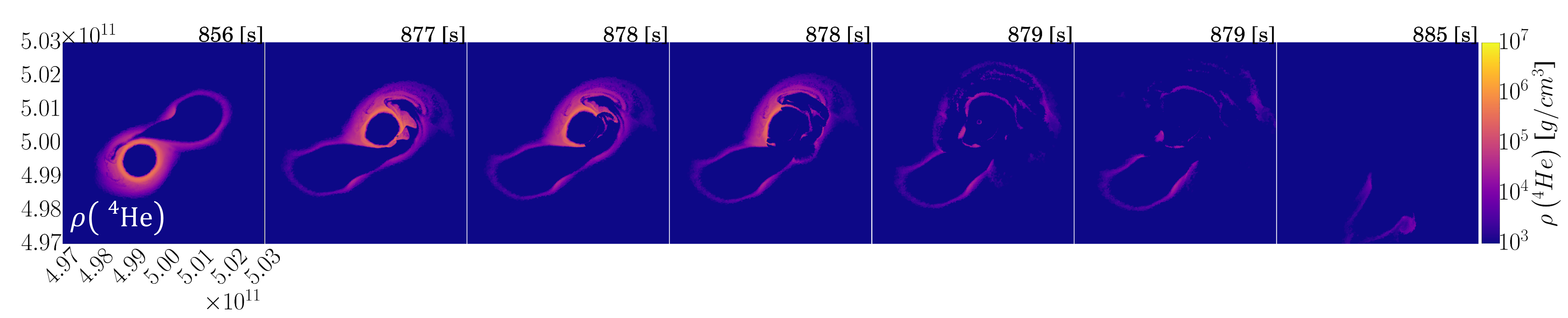}\\
\includegraphics[width=\textwidth]{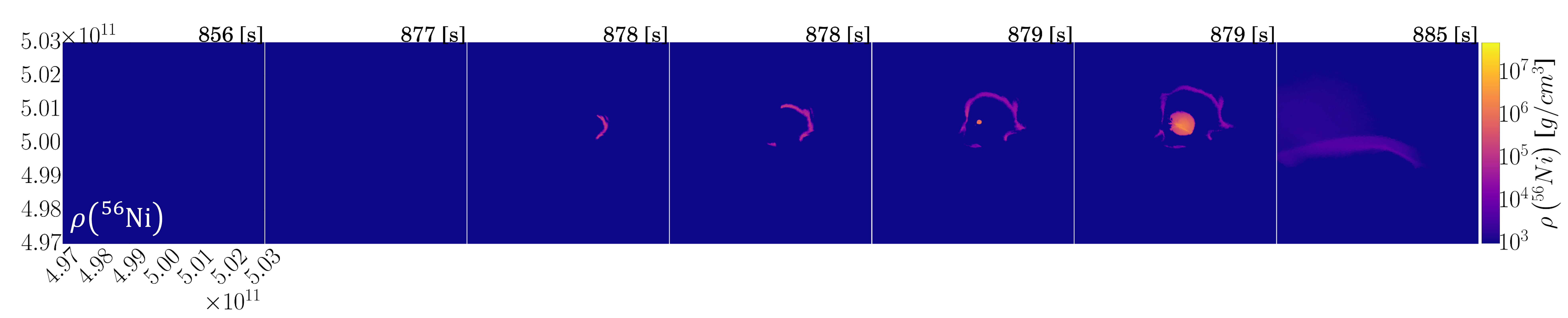}\\
\includegraphics[width=\textwidth]{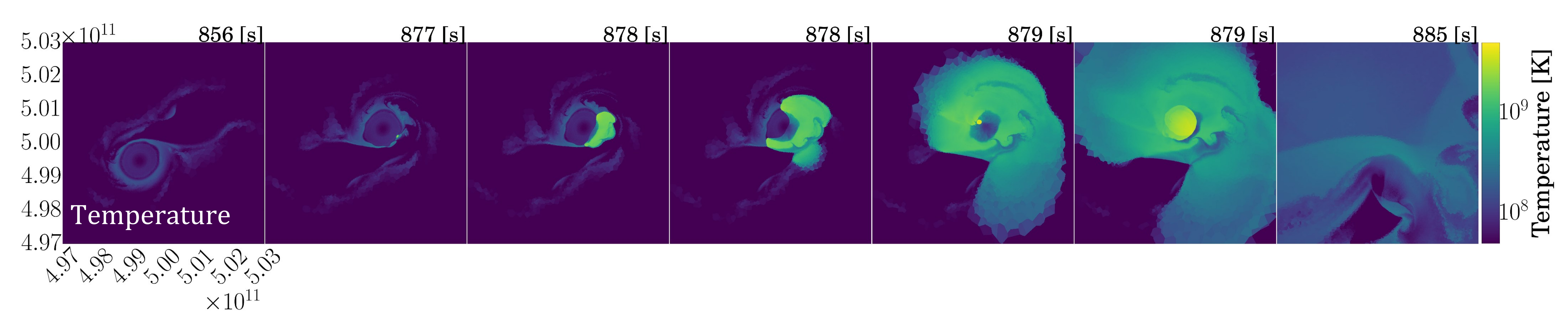}
\caption{\textbf{WD disruption and shock Propagation.} The panels show the time evolution from the time of the partial disruption of the secondary WD (left panels) to the time of the He detonation (second left) when the shock converges in the CO core of the primary WD (third from the left), and finally the ejected remnant in the right panels.}
\label{fig:double_detonation_suplementary}
\end{figure}

In addition, we provide more detailed snapshots of the evolution in \sifig{fig:double_detonation_suplementary}

\begin{figure}
\centering \includegraphics[width=0.5\linewidth,clip]{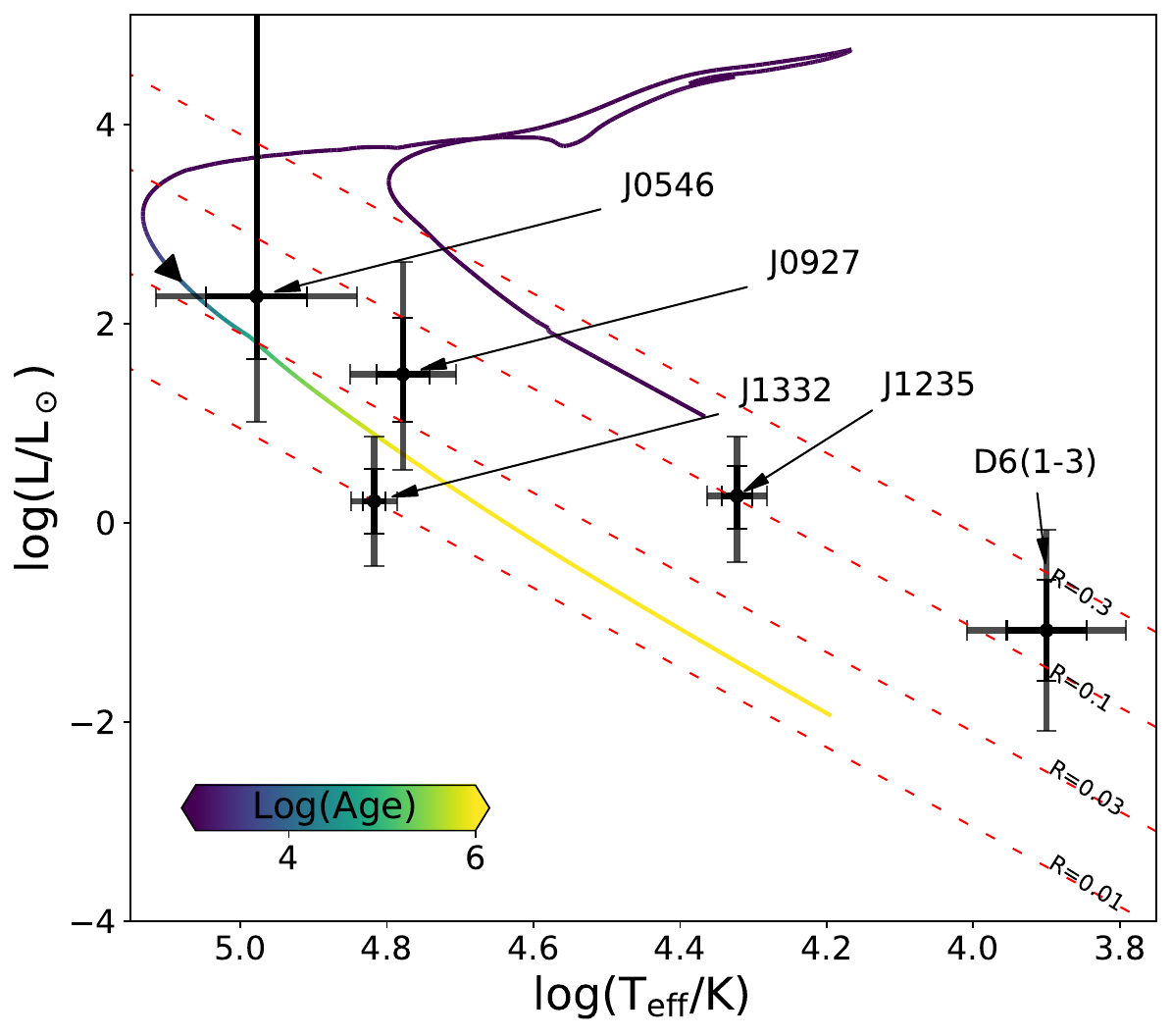}
\includegraphics[width=0.45\linewidth,clip]{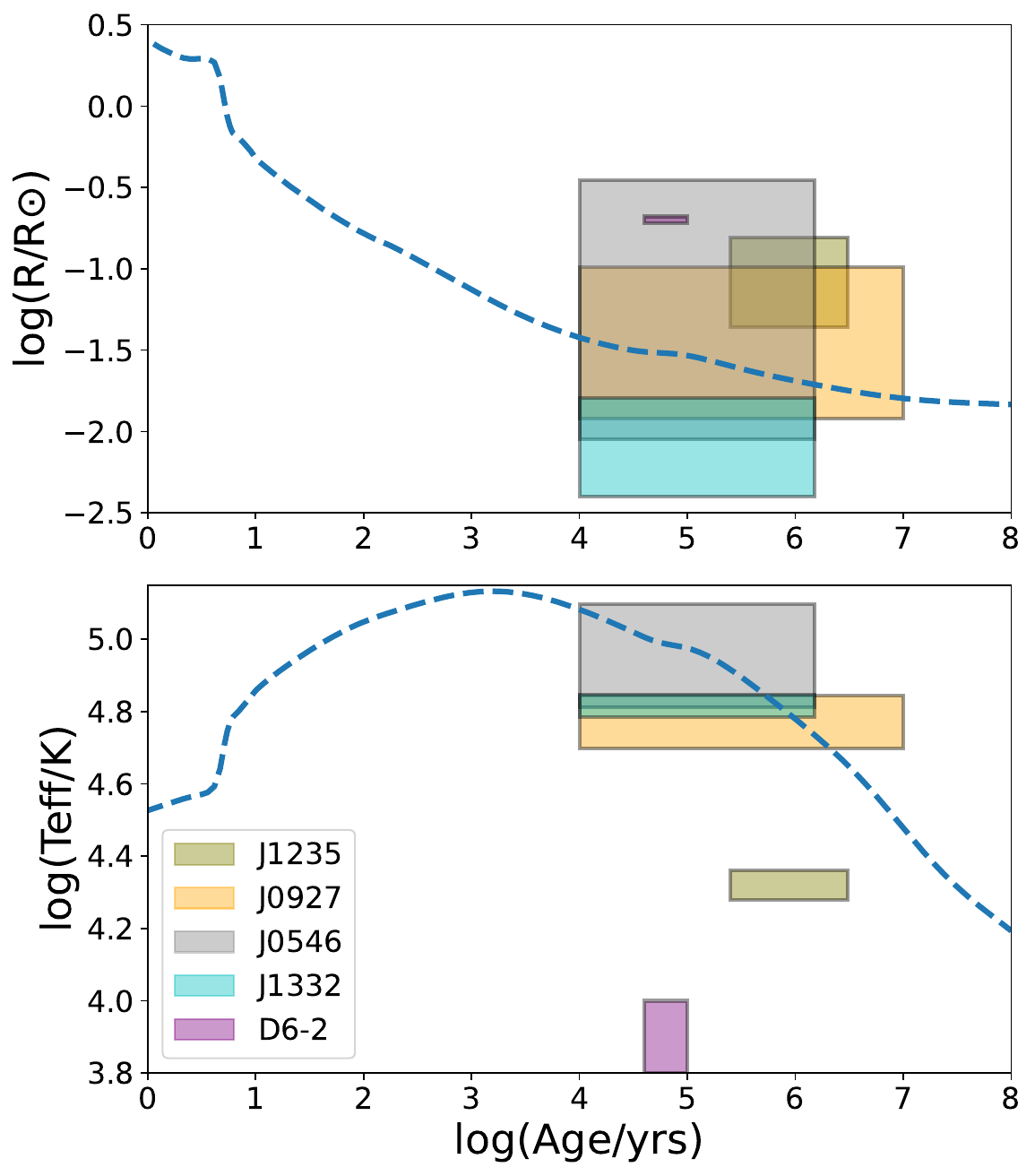}
 \caption{\textbf{Evolution of the MESA model and comparison with all observations. } Left: The HR diagram for our model compared with the observed values, assuming a hydrogen-dominated atmosphere for J1332 based on ref. \cite{Wer+24b}. The error bars depict the 1 and 2 $\sigma$ uncertainties. The black arrow marks the point when the star is $10^4$ years old. Right: The radius and temperature evolution of the model as a function of its age. In particular, the HR positions of HVWDs J0546, J1332, and J0927 may be explained by our model, within the 1-sigma (J0546) and 2-sigma (J0927, J1332); as well as the temperatures of these WDs, while the radius of J1332 is inconsistent, if hydrogen atmosphere is assumed for J1332, but see \ref{sec:SI-J1332}}. 
 \label{fig:HR_hydrogen}
\end{figure}

\subsection{Analytic Estimate of Ejection Velocities}
\label{sec:SI-analytic}
The ejection velocity of the HVWD can be estimated analytically by considering the Keplerian velocity of the binary system at the time of the explosion. Assuming that the explosion occurs near the tidal radius, the ejection velocity of the secondary is given by:

\begin{equation}
v_{\rm ej} \approx \frac{M_1}{M_{\rm tot}}\sqrt{\frac{GM_{\rm tot}}{R_{\rm tid}}},
\end{equation}

where $G$ is the gravitational constant, $M_{\rm tot}$ is the total mass of the binary system, and $R_{\rm tid}$ is the tidal radius. The Tidal radius can be approximated as:

\begin{equation}
R_{\rm tid} \approx R_1 \left(\frac{2M_1}{M_2}\right)^{1/3},
\end{equation}

Where $R_1$ is the radius of the primary WD and $M_1$ and $M_2$ are the masses of the primary and secondary WDs, respectively. The radii of the WDs are calculated using the mass-radius relation formulation by ref. \cite{Nau+72}.

\begin{figure}
\centering
\includegraphics[width=0.6\textwidth]{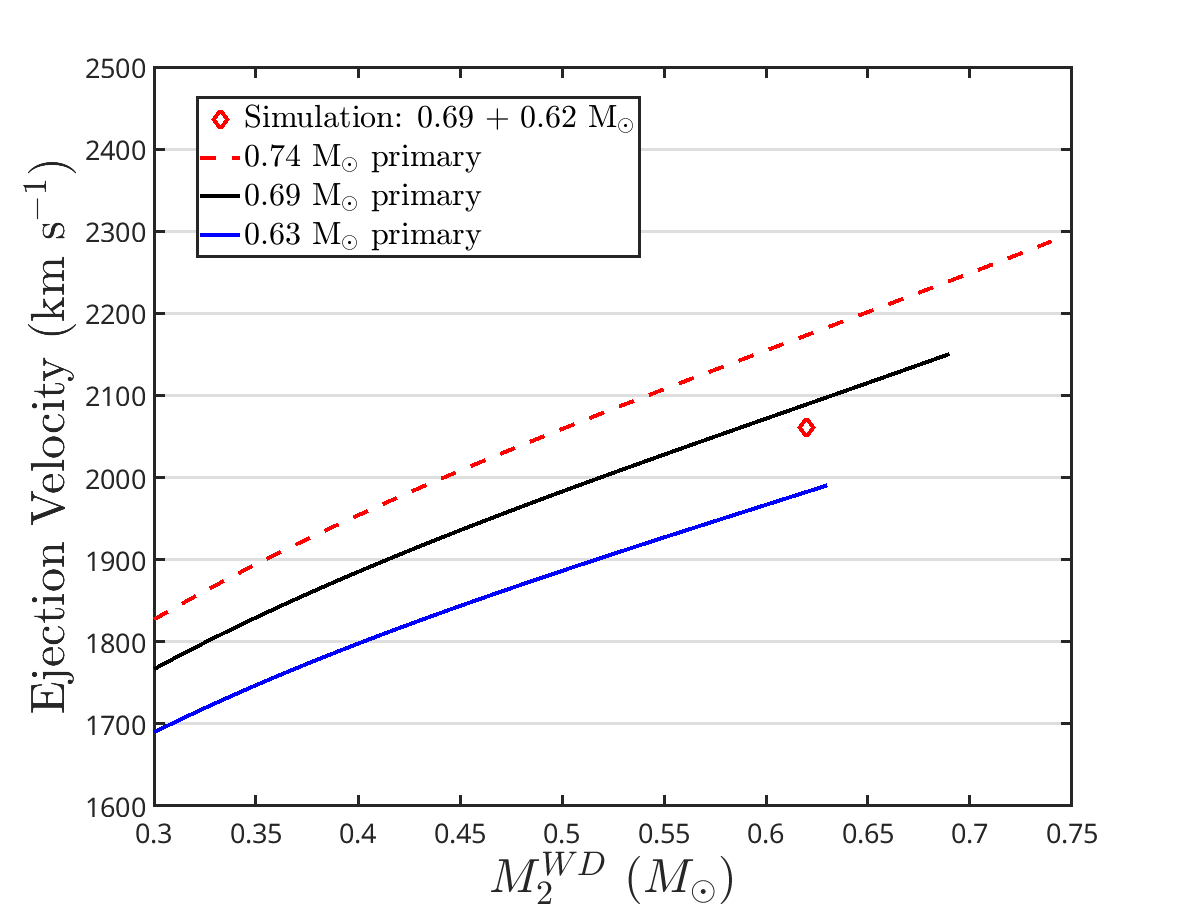}
\caption{\textbf{Expected ejection velocities of partially disrupted WDs from merging binary HeCO WD progenitors.} Three primary masses are shown, corresponding to the upper mass expected for HeCO WDs (0.74 M$_\odot$); the simulated primary (0.69 M$_\odot$); and a lower mass primary expected to experience a double detonation (0.63 M$_\odot$; likely a lower limit). The measured ejection velocity in the simulation well matched the simple analytic estimate.}
\label{fig:velocities}
\end{figure}

\sifig{fig:velocities} shows the expected ejection velocities for a range of binary WD progenitors, calculated using such an analytic estimate. The measured ejection velocity in our simulation matches the analytic prediction well. Nevertheless, this simplified estimate should be taken with a grain of salt, given that the final stages of the tidal evolution and deformation before the SN explosion affect the orbit and the structure of the WDs, and even some of the companion mass is stripped; hence these should only be considered as order of magnitude expectations as to bracket the range of possible ejection velocities in the double HeCO WD merger scenario. 

Both our analytical approximation and the measured companion velocity at the time of the ejection, show them to be comparable with the Keplerian velocity; this suggests that the overall ejection velocity is dominated by the Keplerian velocity. Though the explosion might not be completely symmetric, and somewhat affect the ejection, this effect is limited (and difficult to accurately estimate). We do note that the main detonation occurs close to the center of the WD and is therefore relatively symmetric.

\subsection{The properties of HVWD J1332}
\label{sec:SI-J1332}
Given that in both the D6 scenario and our model, the outer layers of the ejected HVWD are stripped, it is unlikely that such a WD would have a hydrogen atmosphere. Therefore, we do not adapt the properties of the HVWD inferred by \cite{Wer+24b} for J1332 using a hydrogen-rich atmosphere, but the ones assuming a Helium outer atmosphere (hotter and puffier \cite{Elbadry+23}). We note that J1332 might not be a bona-fide HVWD, as its spectroscopic results\cite{Wer+24b} (confirming its hydrogen-rich atmosphere) suggest that its velocity may have been overestimated. Consequently, its total spatial velocity could fall within the typical range for a halo object that is only loosely bound to the Milky Way. Nevertheless, for completeness, in \sifig{fig:HR_hydrogen} we show the HR diagram, radius, and temperature models, and their comparison with observations of J1332 if the hydrogen-rich model of \cite{Wer+24b} is taken at face value. In that case, the HR position is only consistent within $2-\sigma$ and the inferred radius is no longer consistent with our simulation results. One should note that in that case, J1332 could be consistent with a model of a far more massive CO HVWD, as discussed in \cite{Bha+24}.

\subsection{Population Synthesis Considerations}
\label{sec:SI-pop-syn}
Population synthesis studies suggest that mergers involving hybrid WDs are not rare among double WD mergers, with as much as 25\% of WD mergers including at least one hybrid WD \cite{Per+19,Zen+23}. The mergers of double HeCO WDs represent only a fraction of these, with a smaller fraction likely producing the specific conditions required for the partial disruption and double-detonation explosion observed in our simulation. If such explosions are a significant source of HVWDs, their relatively low occurrence rate could explain the paucity of HVWDs compared to the rate of normal Type Ia SNe.

\subsection{Observational Signatures}
The peculiar Type Ia SNe produced by the HeCO WD merger scenario are expected to have fainter luminosities and different observational signatures compared to standard SNe Ia. These differences arise from the lower mass of the exploding WD and the distinct composition of the ejecta. Further studies are needed to characterize the detailed observational properties of these explosions and their potential contribution to the observed diversity of SNe Ia. These will be explored in future studies.


%




\newpage
\begin{thebibliography}{10}
\expandafter\ifx\csname url\endcsname\relax
  \def\url#1{\burl{#1}}\fi
\expandafter\ifx\csname urlprefix\endcsname\relax\def\urlprefix{URL }\fi
\providecommand{\bibinfo}[2]{#2}
\providecommand{\eprint}[2][]{\url{#2}}
\providecommand{\doi}[1]{\url{https://doi.org/#1}}
\bibcommenthead

\bibitem{Jor+12}
\bibinfo{author}{{Jordan}, I., George~C.}, \bibinfo{author}{{Perets}, H.~B.}, \bibinfo{author}{{Fisher}, R.~T.} \& \bibinfo{author}{{van Rossum}, D.~R.}
\newblock \bibinfo{title}{{Failed-detonation Supernovae: Subluminous Low-velocity Ia Supernovae and their Kicked Remnant White Dwarfs with Iron-rich Cores}}.
\newblock \emph{\bibinfo{journal}{\apjl}} \textbf{\bibinfo{volume}{761}}, \bibinfo{pages}{L23} (\bibinfo{year}{2012}).

\bibitem{Ven+17}
\bibinfo{author}{{Vennes}, S.} \emph{et~al.}
\newblock \bibinfo{title}{{An unusual white dwarf star may be a surviving remnant of a subluminous Type Ia supernova}}.
\newblock \emph{\bibinfo{journal}{Science}} \textbf{\bibinfo{volume}{357}}, \bibinfo{pages}{680--683} (\bibinfo{year}{2017}).

\bibitem{Shen2018}
\bibinfo{author}{{Shen}, K.~J.} \emph{et~al.}
\newblock \bibinfo{title}{{Three Hypervelocity White Dwarfs in Gaia DR2: Evidence for Dynamically Driven Double-degenerate Double-detonation Type Ia Supernovae}}.
\newblock \emph{\bibinfo{journal}{\apj}} \textbf{\bibinfo{volume}{865}}, \bibinfo{pages}{15} (\bibinfo{year}{2018}).

\bibitem{Elbadry+23}
\bibinfo{author}{{El-Badry}, K.} \emph{et~al.}
\newblock \bibinfo{title}{{The fastest stars in the Galaxy}}.
\newblock \emph{\bibinfo{journal}{The Open Journal of Astrophysics}} \textbf{\bibinfo{volume}{6}}, \bibinfo{pages}{28} (\bibinfo{year}{2023}).

\bibitem{Rad+18}
\bibinfo{author}{{Raddi}, R.} \emph{et~al.}
\newblock \bibinfo{title}{{Further Insight on the Hypervelocity White Dwarf, LP 40-365 (GD 492): A Nearby Emissary from a Single-degenerate Type Ia Supernova}}.
\newblock \emph{\bibinfo{journal}{\apj}} \textbf{\bibinfo{volume}{858}}, \bibinfo{pages}{3} (\bibinfo{year}{2018}).

\bibitem{Rad+19}
\bibinfo{author}{{Raddi}, R.} \emph{et~al.}
\newblock \bibinfo{title}{{Partly burnt runaway stellar remnants from peculiar thermonuclear supernovae}}.
\newblock \emph{\bibinfo{journal}{\mnras}} \textbf{\bibinfo{volume}{489}}, \bibinfo{pages}{1489--1508} (\bibinfo{year}{2019}).

\bibitem{Gaia2018}
\bibinfo{author}{{Gaia Collaboration}} \emph{et~al.}
\newblock \bibinfo{title}{{Gaia Data Release 2. Summary of the contents and survey properties}}.
\newblock \emph{\bibinfo{journal}{\aap}} \textbf{\bibinfo{volume}{616}}, \bibinfo{pages}{A1} (\bibinfo{year}{2018}).

\bibitem{Gaia2021}
\bibinfo{author}{{Gaia Collaboration}} \emph{et~al.}
\newblock \bibinfo{title}{{Gaia Data Release 3. Summary of the content and survey properties}}.
\newblock \emph{\bibinfo{journal}{\aap}} \textbf{\bibinfo{volume}{674}}, \bibinfo{pages}{A1} (\bibinfo{year}{2023}).

\bibitem{2023IgoshevHypervelocity}
\bibinfo{author}{{Igoshev}, A.~P.}, \bibinfo{author}{{Perets}, H.} \& \bibinfo{author}{{Hallakoun}, N.}
\newblock \bibinfo{title}{{Hyper-runaway and hypervelocity white dwarf candidates in Gaia Data Release 3: Possible remnants from Ia/Iax supernova explosions or dynamical encounters}}.
\newblock \emph{\bibinfo{journal}{\mnras}} \textbf{\bibinfo{volume}{518}}, \bibinfo{pages}{6223--6237} (\bibinfo{year}{2023}).

\bibitem{Sch+24}
\bibinfo{author}{{Scholz}, R.~D.}
\newblock \bibinfo{title}{{Hypervelocity star candidates from Gaia DR2 and DR3 proper motions and parallaxes}}.
\newblock \emph{\bibinfo{journal}{\aap}} \textbf{\bibinfo{volume}{685}}, \bibinfo{pages}{A162} (\bibinfo{year}{2024}).

\bibitem{Bha+24}
\bibinfo{author}{{Bhat}, A.} \emph{et~al.}
\newblock \bibinfo{title}{{Supernova shocks cannot explain the inflated state of hypervelocity runaways from white dwarf binaries}}.
\newblock \emph{\bibinfo{journal}{\aap}} \textbf{\bibinfo{volume}{693}}, \bibinfo{pages}{A114} (\bibinfo{year}{2025}).

\bibitem{Won+24}
\bibinfo{author}{{Wong}, T. L.~S.}, \bibinfo{author}{{White}, C.~J.} \& \bibinfo{author}{{Bildsten}, L.}
\newblock \bibinfo{title}{{Shocking and Mass Loss of Compact Donor Stars in Type Ia Supernovae}}.
\newblock \emph{\bibinfo{journal}{\apj}} \textbf{\bibinfo{volume}{973}}, \bibinfo{pages}{65} (\bibinfo{year}{2024}).

\bibitem{2003HansenTypeIaWithHVWD}
\bibinfo{author}{{Hansen}, B. M.~S.}
\newblock \bibinfo{title}{{Type Ia Supernovae and High-Velocity White Dwarfs}}.
\newblock \emph{\bibinfo{journal}{\apj}} \textbf{\bibinfo{volume}{582}}, \bibinfo{pages}{915--918} (\bibinfo{year}{2003}).

\bibitem{1961BlaauwLick}
\bibinfo{author}{{Blaauw}, A.}
\newblock \bibinfo{title}{{On the origin of the O- and B-type stars with high velocities (the ``run-away'' stars), and some related problems}}.
\newblock \emph{\bibinfo{journal}{\bain}} \textbf{\bibinfo{volume}{15}}, \bibinfo{pages}{265} (\bibinfo{year}{1961}).

\bibitem{Gui+10}
\bibinfo{author}{{Guillochon}, J.}, \bibinfo{author}{{Dan}, M.}, \bibinfo{author}{{Ramirez-Ruiz}, E.} \& \bibinfo{author}{{Rosswog}, S.}
\newblock \bibinfo{title}{{Surface Detonations in Double Degenerate Binary Systems Triggered by Accretion Stream Instabilities}}.
\newblock \emph{\bibinfo{journal}{\apjl}} \textbf{\bibinfo{volume}{709}}, \bibinfo{pages}{L64--L69} (\bibinfo{year}{2010}).

\bibitem{Pakmor2013}
\bibinfo{author}{{Pakmor}, R.}, \bibinfo{author}{{Kromer}, M.}, \bibinfo{author}{{Taubenberger}, S.} \& \bibinfo{author}{{Springel}, V.}
\newblock \bibinfo{title}{{Helium-ignited Violent Mergers as a Unified Model for Normal and Rapidly Declining Type Ia Supernovae}}.
\newblock \emph{\bibinfo{journal}{\apjl}} \textbf{\bibinfo{volume}{770}}, \bibinfo{pages}{L8} (\bibinfo{year}{2013}).

\bibitem{Sato+15}
\bibinfo{author}{{Sato}, Y.} \emph{et~al.}
\newblock \bibinfo{title}{{A Systematic Study of Carbon-Oxygen White Dwarf Mergers: Mass Combinations for Type Ia Supernovae}}.
\newblock \emph{\bibinfo{journal}{\apj}} \textbf{\bibinfo{volume}{807}}, \bibinfo{pages}{105} (\bibinfo{year}{2015}).

\bibitem{Ibe85}
\bibinfo{author}{{Iben}, I., Jr.} \& \bibinfo{author}{{Tutukov}, A.~V.}
\newblock \bibinfo{title}{{On the evolution of close binaries with components of initial mass between 3 solar masses and 12 solar masses}}.
\newblock \emph{\bibinfo{journal}{\apjs}} \textbf{\bibinfo{volume}{58}}, \bibinfo{pages}{661--710} (\bibinfo{year}{1985}).

\bibitem{Iben+87}
\bibinfo{author}{{Iben}, J., Icko}, \bibinfo{author}{{Nomoto}, K.}, \bibinfo{author}{{Tornambe}, A.} \& \bibinfo{author}{{Tutukov}, A.~V.}
\newblock \bibinfo{title}{{On Interacting Helium Star--White Dwarf Pairs as Supernova Precursors}}.
\newblock \emph{\bibinfo{journal}{\apj}} \textbf{\bibinfo{volume}{317}}, \bibinfo{pages}{717} (\bibinfo{year}{1987}).

\bibitem{Bau+21}
\bibinfo{author}{{Bauer}, E.~B.}, \bibinfo{author}{{Chandra}, V.}, \bibinfo{author}{{Shen}, K.~J.} \& \bibinfo{author}{{Hermes}, J.~J.}
\newblock \bibinfo{title}{{Masses of White Dwarf Binary Companions to Type Ia Supernovae Measured from Runaway Velocities}}.
\newblock \emph{\bibinfo{journal}{\apjl}} \textbf{\bibinfo{volume}{923}}, \bibinfo{pages}{L34} (\bibinfo{year}{2021}).

\bibitem{bauer2019}
\bibinfo{author}{{Bauer}, E.~B.}, \bibinfo{author}{{White}, C.~J.} \& \bibinfo{author}{{Bildsten}, L.}
\newblock \bibinfo{title}{{Remnants of Subdwarf Helium Donor Stars Ejected from Close Binaries with Thermonuclear Supernovae}}.
\newblock \emph{\bibinfo{journal}{\apj}} \textbf{\bibinfo{volume}{887}}, \bibinfo{pages}{68} (\bibinfo{year}{2019}).

\bibitem{Shi+23}
\bibinfo{author}{{Shields}, J.~V.} \emph{et~al.}
\newblock \bibinfo{title}{{No Surviving SN Ia Companion in SNR 0509-67.5: Stellar Population Characterization and Comparison to Models}}.
\newblock \emph{\bibinfo{journal}{\apjl}} \textbf{\bibinfo{volume}{950}}, \bibinfo{pages}{L10} (\bibinfo{year}{2023}).

\bibitem{Tan+19}
\bibinfo{author}{{Tanikawa}, A.}, \bibinfo{author}{{Nomoto}, K.}, \bibinfo{author}{{Nakasato}, N.} \& \bibinfo{author}{{Maeda}, K.}
\newblock \bibinfo{title}{{Double-detonation Models for Type Ia Supernovae: Trigger of Detonation in Companion White Dwarfs and Signatures of Companions{\textquoteright} Stripped-off Materials}}.
\newblock \emph{\bibinfo{journal}{\apj}} \textbf{\bibinfo{volume}{885}}, \bibinfo{pages}{103} (\bibinfo{year}{2019}).

\bibitem{2021PakmorZenatiHybridIgnition}
\bibinfo{author}{{Pakmor}, R.}, \bibinfo{author}{{Zenati}, Y.}, \bibinfo{author}{{Perets}, H.~B.} \& \bibinfo{author}{{Toonen}, S.}
\newblock \bibinfo{title}{{Thermonuclear explosion of a massive hybrid HeCO white dwarf triggered by a He detonation on a companion}}.
\newblock \emph{\bibinfo{journal}{\mnras}} \textbf{\bibinfo{volume}{503}}, \bibinfo{pages}{4734--4747} (\bibinfo{year}{2021}).

\bibitem{2022Pakmor}
\bibinfo{author}{{Pakmor}, R.} \emph{et~al.}
\newblock \bibinfo{title}{{On the fate of the secondary white dwarf in double-degenerate double-detonation Type Ia supernovae}}.
\newblock \emph{\bibinfo{journal}{\mnras}} \textbf{\bibinfo{volume}{517}}, \bibinfo{pages}{5260--5271} (\bibinfo{year}{2022}).

\bibitem{2024Boos}
\bibinfo{author}{{Boos}, S.~J.}, \bibinfo{author}{{Townsley}, D.~M.} \& \bibinfo{author}{{Shen}, K.~J.}
\newblock \bibinfo{title}{{Type Ia Supernovae Can Arise from the Detonations of Both Stars in a Double Degenerate Binary}}.
\newblock \emph{\bibinfo{journal}{\apj}} \textbf{\bibinfo{volume}{972}}, \bibinfo{pages}{200} (\bibinfo{year}{2024}).

\bibitem{2024Pollin}
\bibinfo{author}{{Pollin}, J.~M.} \emph{et~al.}
\newblock \bibinfo{title}{{On the fate of the secondary white dwarf in double-degenerate double-detonation Type Ia supernovae - II. 3D synthetic observables}}.
\newblock \emph{\bibinfo{journal}{\mnras}} \textbf{\bibinfo{volume}{533}}, \bibinfo{pages}{3036--3052} (\bibinfo{year}{2024}).

\bibitem{Ibe+85}
\bibinfo{author}{{Iben}, J., I.} \& \bibinfo{author}{{Tutukov}, A.~V.}
\newblock \bibinfo{title}{{On the evolution of close binaries with components of initial mass between 3 M and 12 M.}}
\newblock \emph{\bibinfo{journal}{\apjs}} \textbf{\bibinfo{volume}{58}}, \bibinfo{pages}{661--710} (\bibinfo{year}{1985}).

\bibitem{Zenati2019}
\bibinfo{author}{{Zenati}, Y.}, \bibinfo{author}{{Toonen}, S.} \& \bibinfo{author}{{Perets}, H.~B.}
\newblock \bibinfo{title}{{Formation and evolution of hybrid He-CO white dwarfs and their properties}}.
\newblock \emph{\bibinfo{journal}{\mnras}} \textbf{\bibinfo{volume}{482}}, \bibinfo{pages}{1135--1142} (\bibinfo{year}{2019}).

\bibitem{2010SpringelArepo}
\bibinfo{author}{{Springel}, V.}
\newblock \bibinfo{title}{{E pur si muove: Galilean-invariant cosmological hydrodynamical simulations on a moving mesh}}.
\newblock \emph{\bibinfo{journal}{\mnras}} \textbf{\bibinfo{volume}{401}}, \bibinfo{pages}{791--851} (\bibinfo{year}{2010}).

\bibitem{2011Waldman}
\bibinfo{author}{{Waldman}, R.} \emph{et~al.}
\newblock \bibinfo{title}{{Helium Shell Detonations on Low-mass White Dwarfs as a Possible Explanation for SN 2005E}}.
\newblock \emph{\bibinfo{journal}{\apj}} \textbf{\bibinfo{volume}{738}}, \bibinfo{pages}{21} (\bibinfo{year}{2011}).

\bibitem{2012MNRASSim}
\bibinfo{author}{{Sim}, S.~A.} \emph{et~al.}
\newblock \bibinfo{title}{{2D simulations of the double-detonation model for thermonuclear transients from low-mass carbon-oxygen white dwarfs}}.
\newblock \emph{\bibinfo{journal}{\mnras}} \textbf{\bibinfo{volume}{420}}, \bibinfo{pages}{3003--3016} (\bibinfo{year}{2012}).

\bibitem{2023Zenati}
\bibinfo{author}{{Zenati}, Y.} \emph{et~al.}
\newblock \bibinfo{title}{{The Origins of Calcium-rich Supernovae From Disruptions of CO White Dwarfs by Hybrid He-CO White Dwarfs}}.
\newblock \emph{\bibinfo{journal}{\apj}} \textbf{\bibinfo{volume}{944}}, \bibinfo{pages}{22} (\bibinfo{year}{2023}).

\bibitem{2017bookTaubenberger}
\bibinfo{author}{{Taubenberger}, S.}
\newblock \bibinfo{title}{ in \textit{{The Extremes of Thermonuclear Supernovae}}} (eds \bibinfo{editor}{{Alsabti}, A.~W.} \& \bibinfo{editor}{{Murdin}, P.}) \emph{\bibinfo{booktitle}{Handbook of Supernovae}} \bibinfo{pages}{317} (\bibinfo{publisher}{Springer International Publishing AG}, \bibinfo{year}{2017}).

\bibitem{Per+10}
\bibinfo{author}{{Perets}, H.~B.} \emph{et~al.}
\newblock \bibinfo{title}{{A faint type of supernova from a white dwarf with a helium-rich companion}}.
\newblock \emph{\bibinfo{journal}{\nat}} \textbf{\bibinfo{volume}{465}}, \bibinfo{pages}{322--325} (\bibinfo{year}{2010}).

\bibitem{2011MESA}
\bibinfo{author}{{Paxton}, B.} \emph{et~al.}
\newblock \bibinfo{title}{{Modules for Experiments in Stellar Astrophysics (MESA)}}.
\newblock \emph{\bibinfo{journal}{\apjs}} \textbf{\bibinfo{volume}{192}}, \bibinfo{pages}{3} (\bibinfo{year}{2011}).

\bibitem{Wer+24}
\bibinfo{author}{{Werner}, K.}, \bibinfo{author}{{El-Badry}, K.}, \bibinfo{author}{{G{\"a}nsicke}, B.~T.} \& \bibinfo{author}{{Shen}, K.~J.}
\newblock \bibinfo{title}{{Ultraviolet spectroscopy of the supernova Ia hypervelocity runaway white dwarf J0927‑6335}}.
\newblock \emph{\bibinfo{journal}{\aap}} \textbf{\bibinfo{volume}{689}}, \bibinfo{pages}{L6} (\bibinfo{year}{2024}).

\bibitem{Per+19}
\bibinfo{author}{{Perets}, H.~B.}, \bibinfo{author}{{Zenati}, Y.}, \bibinfo{author}{{Toonen}, S.} \& \bibinfo{author}{{Bobrick}, A.}
\newblock \bibinfo{title}{{Normal type Ia supernovae from disruptions of hybrid He-CO white-dwarfs by CO white-dwarfs}}.
\newblock \emph{\bibinfo{journal}{arXiv e-prints}} \bibinfo{pages}{arXiv:1910.07532} (\bibinfo{year}{2019}).

\bibitem{Zen+18}
\bibinfo{author}{Zenati, Y.}, \bibinfo{author}{Toonen, S.} \& \bibinfo{author}{Perets, H.~B.}
\newblock \bibinfo{title}{{Formation and evolution of hybrid He–CO white dwarfs and their properties}}.
\newblock \emph{\bibinfo{journal}{Monthly Notices of the Royal Astronomical Society}} \textbf{\bibinfo{volume}{482}}, \bibinfo{pages}{1135--1142} (\bibinfo{year}{2018}).
\newblock \urlprefix\url{https://doi.org/10.1093/mnras/sty2723}.

\bibitem{Zen+19a}
\bibinfo{author}{{Zenati}, Y.}, \bibinfo{author}{{Perets}, H.~B.} \& \bibinfo{author}{{Toonen}, S.}
\newblock \bibinfo{title}{{Neutron star-white dwarf mergers: early evolution, physical properties, and outcomes}}.
\newblock \emph{\bibinfo{journal}{\mnras}} \textbf{\bibinfo{volume}{486}}, \bibinfo{pages}{1805--1813} (\bibinfo{year}{2019}).

\bibitem{2011PakmorArepoMHD}
\bibinfo{author}{{Pakmor}, R.}, \bibinfo{author}{{Bauer}, A.} \& \bibinfo{author}{{Springel}, V.}
\newblock \bibinfo{title}{{Magnetohydrodynamics on an unstructured moving grid}}.
\newblock \emph{\bibinfo{journal}{\mnras}} \textbf{\bibinfo{volume}{418}}, \bibinfo{pages}{1392--1401} (\bibinfo{year}{2011}).

\bibitem{2015ZhuMHDArepoWD}
\bibinfo{author}{{Zhu}, C.}, \bibinfo{author}{{Pakmor}, R.}, \bibinfo{author}{{van Kerkwijk}, M.~H.} \& \bibinfo{author}{{Chang}, P.}
\newblock \bibinfo{title}{{Magnetized Moving Mesh Merger of a Carbon-Oxygen White Dwarf Binary}}.
\newblock \emph{\bibinfo{journal}{\apjl}} \textbf{\bibinfo{volume}{806}}, \bibinfo{pages}{L1} (\bibinfo{year}{2015}).

\bibitem{2020WeinbergArepoPublic}
\bibinfo{author}{{Weinberger}, R.}, \bibinfo{author}{{Springel}, V.} \& \bibinfo{author}{{Pakmor}, R.}
\newblock \bibinfo{title}{{The AREPO Public Code Release}}.
\newblock \emph{\bibinfo{journal}{\apjs}} \textbf{\bibinfo{volume}{248}}, \bibinfo{pages}{32} (\bibinfo{year}{2020}).

\bibitem{Paxton2013}
\bibinfo{author}{{Paxton}, B.} \emph{et~al.}
\newblock \bibinfo{title}{{Modules for Experiments in Stellar Astrophysics (MESA): Planets, Oscillations, Rotation, and Massive Stars}}.
\newblock \emph{\bibinfo{journal}{\apjs}} \textbf{\bibinfo{volume}{208}}, \bibinfo{pages}{4} (\bibinfo{year}{2013}).

\bibitem{Paxton2019}
\bibinfo{author}{{Paxton}, B.} \emph{et~al.}
\newblock \bibinfo{title}{{Modules for Experiments in Stellar Astrophysics (MESA): Pulsating Variable Stars, Rotation, Convective Boundaries, and Energy Conservation}}.
\newblock \emph{\bibinfo{journal}{\apjs}} \textbf{\bibinfo{volume}{243}}, \bibinfo{pages}{10} (\bibinfo{year}{2019}).

\bibitem{Glanz2025Zenodo}
\bibinfo{author}{Bhat, A.}
\newblock \bibinfo{title}{Mesa files for origins of the fastest stars from merger-disruption of he-co white dwarfs} (\bibinfo{year}{2025}).
\newblock \urlprefix\url{https://doi.org/10.5281/zenodo.15700950}.

\bibitem{pak+16}
\bibinfo{author}{{Pakmor}, R.} \emph{et~al.}
\newblock \bibinfo{title}{{Improving the convergence properties of the moving-mesh code AREPO}}.
\newblock \emph{\bibinfo{journal}{\mnras}} \textbf{\bibinfo{volume}{455}}, \bibinfo{pages}{1134--1143} (\bibinfo{year}{2016}).

\bibitem{2Jer+23}
\bibinfo{author}{{Jermyn}, A.~S.} \emph{et~al.}
\newblock \bibinfo{title}{{Modules for Experiments in Stellar Astrophysics (MESA): Time-dependent Convection, Energy Conservation, Automatic Differentiation, and Infrastructure}}.
\newblock \emph{\bibinfo{journal}{\apjs}} \textbf{\bibinfo{volume}{265}}, \bibinfo{pages}{15} (\bibinfo{year}{2023}).

\bibitem{Hunter:2007}
\bibinfo{author}{Hunter, J.~D.}
\newblock \bibinfo{title}{Matplotlib: A 2d graphics environment}.
\newblock \emph{\bibinfo{journal}{Computing in Science \& Engineering}} \textbf{\bibinfo{volume}{9}}, \bibinfo{pages}{90--95} (\bibinfo{year}{2007}).

\bibitem{harris2020arrayNUMPY}
\bibinfo{author}{Harris, C.~R.} \emph{et~al.}
\newblock \bibinfo{title}{Array programming with {NumPy}}.
\newblock \emph{\bibinfo{journal}{Nature}} \textbf{\bibinfo{volume}{585}}, \bibinfo{pages}{357--362} (\bibinfo{year}{2020}).
\newblock \urlprefix\url{https://doi.org/10.1038/s41586-020-2649-2}.

\bibitem{2020SciPy-NMeth}
\bibinfo{author}{Virtanen, P.} \emph{et~al.}
\newblock \bibinfo{title}{{{SciPy} 1.0: Fundamental Algorithms for Scientific Computing in Python}}.
\newblock \emph{\bibinfo{journal}{Nature Methods}} \textbf{\bibinfo{volume}{17}}, \bibinfo{pages}{261--272} (\bibinfo{year}{2020}).

\bibitem{Cha+22}
\bibinfo{author}{{Chandra}, V.} \emph{et~al.}
\newblock \bibinfo{title}{{The SN Ia runaway LP 398-9: detection of circumstellar material and surface rotation}}.
\newblock \emph{\bibinfo{journal}{\mnras}} \textbf{\bibinfo{volume}{512}}, \bibinfo{pages}{6122--6133} (\bibinfo{year}{2022}).

\bibitem{Wer+24b}
\bibinfo{author}{{Werner}, K.}, \bibinfo{author}{{Reindl}, N.}, \bibinfo{author}{{Rauch}, T.}, \bibinfo{author}{{El-Badry}, K.} \& \bibinfo{author}{{B{\'e}dard}, A.}
\newblock \bibinfo{title}{{The photospheres of the hottest fastest stars in the Galaxy}}.
\newblock \emph{\bibinfo{journal}{\aap}} \textbf{\bibinfo{volume}{682}}, \bibinfo{pages}{A42} (\bibinfo{year}{2024}).

\bibitem{Nau+72}
\bibinfo{author}{{Nauenberg}, M.}
\newblock \bibinfo{title}{{Analytic Approximations to the Mass-Radius Relation and Energy of Zero-Temperature Stars}}.
\newblock \emph{\bibinfo{journal}{\apj}} \textbf{\bibinfo{volume}{175}}, \bibinfo{pages}{417} (\bibinfo{year}{1972}).

\bibitem{Zen+23}
\bibinfo{author}{{Zenati}, Y.} \emph{et~al.}
\newblock \bibinfo{title}{{The Origins of Calcium-rich Supernovae From Disruptions of CO White Dwarfs by Hybrid He-CO White Dwarfs}}.
\newblock \emph{\bibinfo{journal}{\apj}} \textbf{\bibinfo{volume}{944}}, \bibinfo{pages}{22} (\bibinfo{year}{2023}).

\end{thebibliography}
\end{document}